\documentclass[twocolumn]{aastex631}
\usepackage{amsmath}
\usepackage{comment}

\submitjournal{ApJ}

\shorttitle{High-velocity outflows in [O{\small III}] emitters at $2.5 < z < 9$}
\shortauthors{Cooper et al.}
\begin{document}

\title{High-velocity outflows in [O{\small III}] emitters at $2.5 < z < 9$ from JWST NIRSpec medium-resolution spectroscopy}

\correspondingauthor{Ryan A. Cooper}
\affiliation{Kapteyn Astronomical Institute, University of Groningen, P.O. Box 800, 9700AV Groningen, The Netherlands}
\email{cooper@astro.rug.nl}
\author[0009-0000-0413-5699]{Ryan A. Cooper}
\affiliation{Kapteyn Astronomical Institute, University of Groningen, P.O. Box 800, 9700AV Groningen, The Netherlands}

\author[0000-0001-8183-1460]{Karina I. Caputi}
\affiliation{Kapteyn Astronomical Institute, University of Groningen, P.O. Box 800, 9700AV Groningen, The Netherlands}
\affiliation{Cosmic Dawn Center (DAWN), Copenhagen, Denmark
}

\author[0000-0001-8386-3546]{Edoardo Iani}
\affiliation{Kapteyn Astronomical Institute, University of Groningen, P.O. Box 800, 9700AV Groningen, The Netherlands}
\affiliation{Present address: Institute of Science and Technology Austria (ISTA), Am Campus 1, 3400 Klosterneuburg, Austria}

\author[0000-0002-5104-8245]{Pierluigi Rinaldi}
\affiliation{Steward Observatory, University of Arizona, 933 North Cherry Avenue, Tucson, AZ 85721, USA}

\author[0000-0001-8325-1742]{Guillaume Desprez}
\affiliation{Kapteyn Astronomical Institute, University of Groningen, P.O. Box 800, 9700AV Groningen, The Netherlands}

\author[0000-0001-6066-4624]{Rafael Navarro-Carrera}
\affiliation{Kapteyn Astronomical Institute, University of Groningen, P.O. Box 800, 9700AV Groningen, The Netherlands}

\begin{abstract}

We identify galaxies hosting ionised-gas high-velocity outflows from the complete sample of medium resolution (R1000) JWST/NIRSpec MSA spectroscopy taken as part of the JWST Advanced Deep Extragalactic Survey (JADES). From a total sample of 1087 [O{\small III}]$_{5007}$~emitters we identify 40 galaxies with a blue-side broadened [O{\small III}]$_{5007}$~line at $2.5<z<9$. Of these, 34 are strong outflow candidates whilst 6 sources have broadening potentially driven by rotating clumps. Our outflow candidate sample is mainly composed of star-forming galaxies, including $\sim$65\% starbursts,  which span the stellar mass range $7.5<log_{10}(M_{*}/M_{\odot})<11.0$. It also includes two candidate type-2 active galactic nuclei (AGN) and a `little red dot' (LRD). We report a median outflow velocity of 531$^{+146}_{-159}$~km~s$^{-1}$ and an overall incidence rate of 3.4\%. These values are significantly higher and lower respectively than recent similar works, which we accredit to the limiting resolution of the R1000 spectroscopy and a stricter outflow selection criterion. We find no correlation between the outflow velocity and the galaxy stellar mass or star-formation rate. The median ratio between outflow velocity and escape velocity is 0.77$^{+0.36}_{-0.32}$, indicating that most outflows cannot escape the galaxy gravitational potentials. We do find an anti-correlation between mass loading factor and stellar mass up to $M_\ast \approx 10^{10} \, \rm M_\odot$, with most lowest stellar-mass ($M_\ast <  10^{9} \, \rm M_\odot$) galaxies reaching values well above unity, as is the case for local starburst galaxies.

\end{abstract}

\keywords{High-redshift galaxies (734) --- Galaxy winds(626) --- Galaxy evolution(594) --- Infrared astronomy(786)}

\section{Introduction} \label{sec:intro}

Understanding the processes that regulate star formation in galaxies is imperative to building a complete picture of galaxy evolution. Critically, the rate of star formation in galaxies is typically of the order of a few percent of available baryonic mass across all redshifts (e.g.,  \citealt{sfe-z-1,sfe_z-1-2} at $0 < z < 2$, \citealt{m_star_mh_2,sfe-z-to-5-2, sfe-z-to-5-1} out to $z\sim5$; and \citealt{sfe-high-z,sfe-higher-z} out to $z\sim10$), far lower than the expected 50-80\% predicted by cosmological models when feedback is excluded \citep{high-sfe-sims-1,high-sfe-sims-2,high-sfe-sims-3}. 

These regulation mechanisms are the likely drivers of the evolution of galaxies towards a passive, quiescent state dominated by evolved stellar populations. This quiescent phase can be temporary, referred to as `mini-quenching' \citep{10.1093/mnras/stad3239, looser2024recentlyquenchedgalaxy700}, or more permanent, leading to the classical definition of passive galaxies. These passive galaxies are abundant in the local Universe \citep{quiescent_2004,quiescent-local-2, quiescent_2007,quiescent-local-1, quiescent-local-3} and have been observed up to intermediate redshifts \citep[$1 \lesssim$ z $\lesssim 3$ e.g., ][]{quiescent-int-1,queiscent-int-2,quiescent-int-3},  where quenching has had significant time to halt star formation. In addition, recent observations have also identified higher redshift (z $\geq$ 3) galaxies with the low star formation rates (SFR) and stellar ages associated with some local ellipticals \citep{quiescent-high-z-1,quiescent-high-z-4,quiescent-high-z-3, quiescent-high-z-2}. For such galaxies to form, feedback must be present and provide significant quenching at substantially high redshifts. 

Whilst the precise significance of varying feedback mechanisms and their impact on galaxy evolution is subject to intense research, outflowing gas driven by stellar and AGN feedback is thought of as a primary contributor (see \citealt{outflow-imps} for a detailed discussion). These outflows have been studied extensively in the local Universe out to beyond cosmic noon ($0<z<3$). A bulk of these works rely on the observation of gas dynamics traced by emission lines in integral field spectroscopy \citep{IFS-outflow-2, IFS-outflow-1, IFS-outflow-3, IFS-outflow-4}, but wider statistical studies have focused on observing the broadening of the luminous H$\alpha$ and [O{\small III}] emission from spatially integrated spectra (e.g., \citealt{int-outflow-3, int-outflow-1, int-outflow-2, int-outflow-4}). 

These observations show that a significant percentage of galaxies support outflowing gas, with percentages rising to $\sim$65\% in high mass samples \citep{int-outflow-4} but a more typical incidence approaches $\sim 30\%$ \citep[e.g., ][]{broad-comp-dust,out-indicence-2}. Observed outflows have a diverse range of velocities, with those driven by AGN reaching the order of a thousand km~s$^{-1}$ whilst stellar feedback produces outflows on the order of 100$^{2}$ km~s$^{-1}$ \citep{outflow-v-1}. The redshift evolution of such outflows is currently not well constrained and a complete description of the impact that outflowing gas has on galaxy evolution requires the study of significant galaxy samples across cosmic time.

To this end, recent studies using JWST spectroscopy have endeavored to trace galactic outflows at previously unexplored redshifts (\citealt{xu2023stellaragnfeedbackprobed} hereafter Xu23; \citealt{2024ApJ...970...19Z}, \citealt{2024A&A...685A..99C} hereafter Ca24). These studies report a diverse range of outflow velocities, with no evolution in redshift, and discuss a variety of scenarios in which the outflowing gas affects the host galaxy. These works act as a successful template for further outflow studies at high redshift with JWST, but larger statistical samples are required to draw significant conclusions on the importance of outflows on galaxy evolution through cosmic time.

With this in mind, we aim with this work to exploit medium resolution JWST NIRSpec spectra to identify and characterise high-velocity, star-formation driven outflows of ionised gas in the high redshift ($2.5 < z < 9$) and  mostly at low stellar masses ($< 10^{10} \, \rm M_{\odot}$) regime. Particularly, we focus our study on the search of asymmetric, blue-shifted broadening in the [O{\small III}]$_{5007}$~emission line. This approach allows us to identify the most robust warm-gas outflow candidates, excluding cases where the line broadening is simply due to high dispersion in the gas velocity field.

For our work we make use of the rich photometric and spectroscopic data available in the Great Observatories Origins Deep Survey (GOODS, \citealt{GOODS, Giavalisco_2004}) fields to determine the physical properties of our outflow candidates and evaluate them against a statistical sample of line emitters across the same redshift range. We explore a significantly larger sample ($>$1000 galaxies) than that of previous high redshift studies to examine the statistical properties of galaxy outflows whilst discussing the feasibility of medium resolution spectroscopy for this purpose. This in turn will inform future studies specifically designed to identify outflows.

This paper is organised as follows. In section \ref{sec:Data} we outline the spectroscopic and photometric data used in this study and in section \ref{sec:Ident} we discuss our sample selection and outflow identification criteria. Section \ref{sec:props} describes the physical parameters and morphologies of our outflow candidates and in section \ref{sec:discussion} we discuss possible selection effects, sources of outflows and the relationship between these outflows and their hosts. We conclude our findings in \ref{sec:conc} and offer thoughts for future outflow investigations.

Throughout this work we adopt the cosmological parameters described in \cite{planck} with values H$_{0}=67.7$km s$^{-1}$Mpc$^{-1}$, $\Omega_{m}=0.308$ and $\Omega_{\Lambda}=0.7$. 

\section{Data} \label{sec:Data}

For this work, we make use of the JWST Advanced Deep Extragalactic Survey (JADES, \citealt{2023arXiv230602465E, 2023ApJS..269...16R}, DOI: \href{https://archive.stsci.edu/hlsp/jades}{10.17909/8tdj-8n28}) spectroscopy available in the GOODS-N and GOODS-S fields. The publicly available Data Release 3 \citep{2024arXiv240406531D} incorporates JWST NIRSpec MSA spectroscopy for many thousands of galaxies, with the majority of these observed with both the low resolution clear prism and three medium resolution gratings. To support these spectra, a wealth of photometry produced by JWST NIRCam and the Hubble Space Telescope (HST) covers both fields. 

Here we outline the data used in this study and provide a broad description of any additional data reduction steps.

\subsection{JWST NIRSpec MSA Spectroscopy} \label{sec:MSA}

We incorporate into our study the complete spectroscopic observations from the JADES program covering GOODS-S (PIDs: 1180, 1210, 1286, 3215 PIs: Eisenstein, Luetzgendorf) and GOODS-N (PID: 1181, PI: Eisenstein); see \citet{Bunker2024}. For each object, low resolution (R $\sim$ 100) clear, prism observations provide spectral coverage from 0.6-5.3$\mu$m whilst three medium resolution (R $\sim$ 1000) gratings (G140M, G235M and G395M) provide higher resolution coverage at wavelength ranges 0.97–1.84$\mu$m, 1.66–3.07$\mu$m and 2.87–5.10$\mu$m respectively. The complete reduction process for these spectra is detailed in \cite{2024arXiv240406531D}, though we provide a brief technical description here.

Each NIRSpec MSA target was observed using two dithers and the 3-point nodding pattern. For the prism spectroscopy a total of 12 integrations produce 12ks observations which has produced robust, high S/N spectra that contain a wealth of emission lines up to z-14 \citep{gnz-14} as well as visible continua in many cases. The medium resolution gratings consist of 6 integrations that result in 6ks observations. These spectra lack the sensitivity of the prism observations and the galaxies in our study frequently show no continua, but bright emission lines are still prominent and observed with robust S/N.  

\begin{figure}
    \centering
    \includegraphics[width=1\linewidth]{ 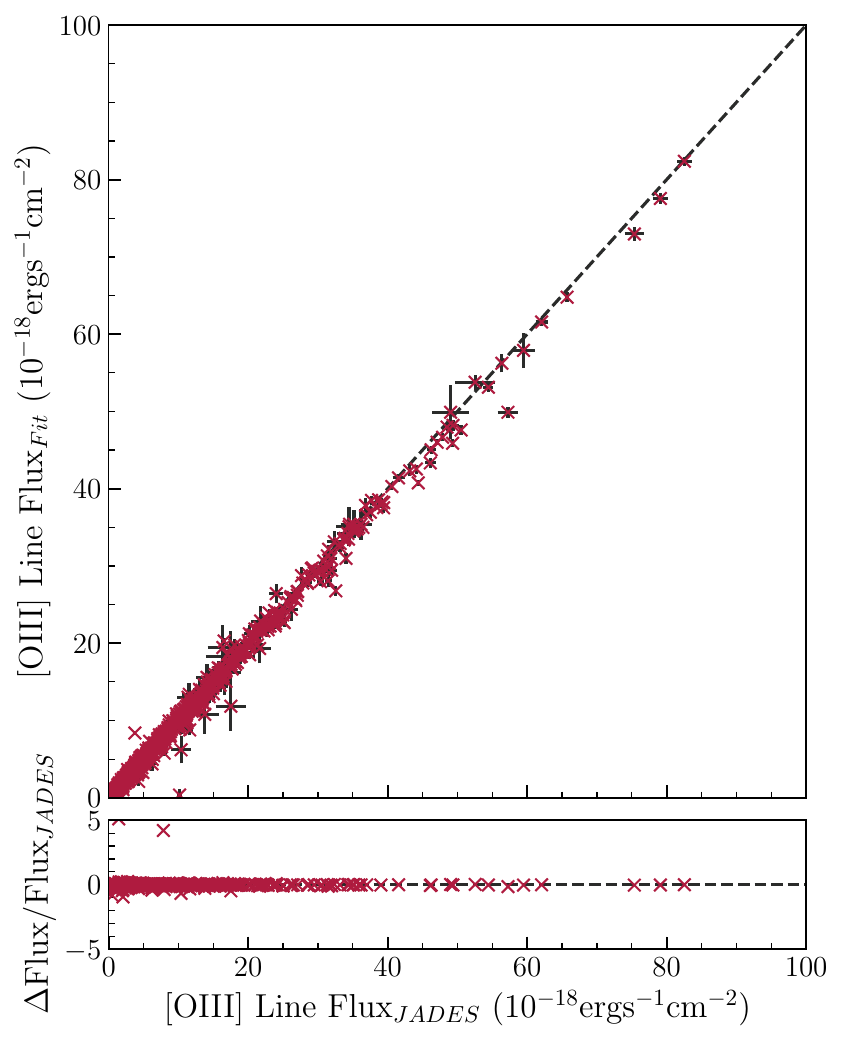}
    \caption{[O{\small III}] line fluxes from the JADES public data release \citep{2023arXiv230602465E, 2023ApJS..269...16R}compared to our line fluxes extracted with MSAEXP.}\label{fig:flux_calib}
    \label{fig:enter-label}
\end{figure}

The bulk of the spectroscopic/spectro-photometric properties obtained for our sample are calculated with the prism spectroscopy. The greater sensitivity allows for a more robust detection of faint emission lines and the presence of continua permits a more accurate calculation of equivalent widths. We perform our line fitting using \texttt{MSAEXP} \citep{brammer_2023_8319596}, but limit this process to the prism spectra for the reasons described prior. \texttt{MSAEXP} uses a combination of broad and narrow templates for emission lines and models the continuum with a composite of splines. In Fig. \ref{fig:flux_calib} we compare our line fluxes to those reported in the publicly available line flux catalogs produced by JADES and find exceptional agreement within errors, with 93(81)\%  of objects within a 10(5)\% difference. 

We finally note that, in some cases, there is a non-negligible shift in the spectral axis between the prism spectroscopy and each medium grating. This results in a change in spectroscopic redshift dependent on the filter/grating combination used. To remain consistent throughout this study and with other works, the spectroscopic redshifts reported in this study are therefore those calculated by our line fitting on the prism spectra. This fitting procedure takes the redshift provided by the JADES data release as a prior with room to vary by $\pm$0.1. Ultimately, all objects produce a robust and appropriate fit within the defined redshift range. 

\subsection{HST and JWST NIRCam Photometry}

The GOODS fields are covered by an extensive array of multiwavelength observations. For our work, we make use of JWST imaging collected by the JADES, JEMS \citep{2023ApJS..268...64W} and FRESCO \citep{2023MNRAS.525.2864O} programs, as well as the wealth of legacy observations with HST \citep{Giavalisco_2004}. 

JADES is one of the largest Cycle 1 JWST programs, consisting of over 700 hours of observations across three of JWST's instruments. Both GOODS fields were observed in 9 NIRCam filters with photometric coverage ranging between $\sim$0.9-5.0$\mu$m. The GOODS-S observations consist of 8 filters with exposure time 12368s and 1 filter (F115W) with exposure time 24737s. This results in a median 5$\sigma$ (0$^{\prime\prime}$.3 aperture) depth across all filters of AB mag $\sim$ 29.74 \citep{hainline2024cosmosinfancyjadesgalaxy}. The GOODS-N field was observed with the same filters with exposures varying from 8503s in the medium bands F335M and F410M to 22676s in F115W resulting in a median 5$\sigma$ depth of AB mag $\sim$ 29.53 measured using a 0$^{\prime\prime}$.3 aperture \citep{2024arXiv240406531D}. 

The JEMS program (PID: 1963, PI: Williams, DOI: \href{https://archive.stsci.edu/hlsp/jems}{10.17909/fsc4-dt61}) is a small Cycle 1 JWST program that covers a total area of 15.6 square arcminutes in GOODS-S. A total of 20.5 hours is split across 5 medium band filters, resulting in exposure times between 13915-27830s for each filter. These observations provide an average 5$\sigma$ depth of $\sim$28.8 AB mag measured with a 0$^{\prime\prime}$.3 aperture across two spectral ranges, the shorter wavelength range $\sim$1.7-2.2$\mu$m and the longer wavelength range $\sim$4.2-5.0$\mu$m \citep{2023ApJS..268...64W}.

FRESCO (PID: 1895, PI: Oesch, DOI: \href{https://archive.stsci.edu/hlsp/fresco}{10.17909/gdyc-7g80}) is a medium size Cycle 1 JWST program that focused on obtaining NIRCam/WFSS spectroscopy in both GOODS fields. As well as this spectroscopy, the $\sim$54 hour program obtains NIRCam imaging in the short wavelength F182M and F210M filters as well as the long wavelength F444W filter. The 5$\sigma$ depth of each filter are 28.2, 28.2 and 28.4 AB mag respectively measured with 0$^{\prime\prime}$.32 apertures \citep{2023MNRAS.525.2864O}.

We make use of the publicly available photometric catalogs released by the JADES team in data release 3. The catalogs include photometry from all the programs described above as well as substantial legacy photometry from HST \citep{2019ApJS..244...16W}. We refer the reader to \cite{2023NatAs...7..611R} and \cite{Tacchella_2023} for descriptions of the photometric analysis and proceed to use Kron aperture derived photometry for this study.

\section{Identification of galaxy outflows} \label{sec:Ident}

We analyse the complete JADES spectroscopic sample to build a selection of candidate galaxies exhibiting significant outflows. Here we outline our purely spectroscopic approach to identifying candidates and the steps taken to discount spurious sources, focusing on the limitations of the medium resolution spectra and the necessary steps to account for instrumental broadening.

\subsection{Initial selection of [O{\small III}] emitters} \label{sec:InitialSelect}

In this work, we identify outflow candidates from the analysis of the [O{\small III}]$_{5007}$~emission line. This line has the benefit of being sufficiently bright for our needs whilst still tracing gas within the galaxy. By limiting our initial approach to [O{\small III}] emission only, we reduce the likelihood of our candidate outflows being a result of broad-line AGN activity. 

As discussed in section \ref{sec:MSA}, we perform line fitting on each prism spectrum using \texttt{MSAEXP}. We set a redshift prior for each source centered on the spectroscopic redshift determined in the JADES DR3 line flux catalogs, but allow this value to vary by $\pm$0.1 to account for uncertainties in the fitting process. In this way, we obtain redshifts, line fluxes and equivalent widths for every source in the JADES observations. We then collect every object with a measured [O{\small III}] detection and perform an initial redshift cut, removing objects with spec-z $<$ 2.5. With this cut we are able to investigate any evolutionary trend in outflows and their hosts from near cosmic noon out to the earliest epochs. This produces a total initial sample of 1227 galaxies spanning a redshift range of 2.5 $<$ z $<$ 9.1.

With this sample, we introduce two further selection criteria. These criteria are deliberately broad and less stringent than those in Xu23 and Ca24, in order to increase the number of galaxies in our final sample and avoid artificially excluding potential sources.  The first of these criteria is a minimum rest-frame equivalent width (EW$_{0}$) of 100\AA \space for the [O{\small III}]$_{5007}$~line. At anything but the highest redshifts (z $\gtrsim$ 5.0) the [O{\small III}] doublet is sufficiently or partly blended in the PRISM data to the point where each line cannot be reliably separated. In these cases we choose to include objects whose combined [O{\small III}] EW$_{0}$ is $>$ 100\AA \space with the knowledge that the two lines will be resolved in the medium resolution gratings. Note that sources for which the continuum in the PRISM spectra is not detected are directly discarded from the sample. In practice, we do not expect this to remove any possible candidates as the medium resolution spectra are considerably less sensitive and these objects would likely not have robust enough spectra to identify outflowing components anyway.

\begin{figure}
    \centering
    \includegraphics[width=1\linewidth]{ 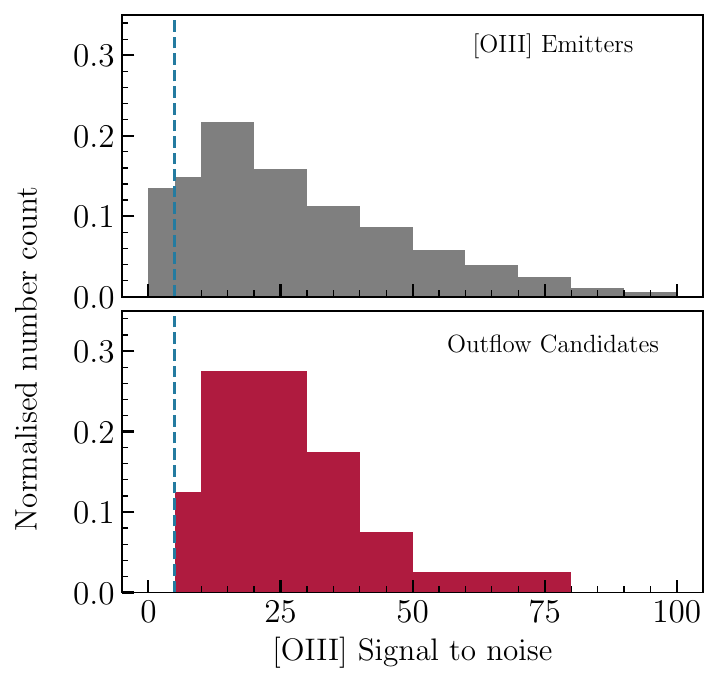}
    \caption{The normalised distribution of [O{\small III}] emitters (grey) and our final outflow candidates (red). The dashed blue line at S/N $>$ 5 is the lower limit at which we first identify outflows within our sample.}
    \label{fig:snr}
\end{figure}

We secondly impose an initial signal-to-noise (S/N) ratio cut on the [O{\small III}]$_{5007}$~line of S/N $>$ 3. We measure the S/N by summing the flux of the emission line and the associated error across the same line region. Previous studies have identified potential biases in the statistical distribution of outflows and the relationships between outflows and their hosts physical parameters. Namely, a higher S/N, for a given instrument sensitivity, suggests a brighter line indicative of enhanced star formation \citep{sfr-vs-oiii-1, sfr-vs-oiii-2}. Indeed, recent, high redshift, outflow studies have found correlations with outflow incidence and SFR \citep{xu2023stellaragnfeedbackprobed, 2024A&A...685A..99C}, but care has to be taken that this is not a consequence of S/N.  The full S/N distribution of our [O{\small III}] emitting sample as well as the selected outflow candidates is shown in Fig. \ref{fig:snr}. We see that a S/N cut of 3 is effective at obtaining the faintest [O{\small III}] emitters, but note that no outflows are detected at this S/N level. In fact, we do not observe any outflows until S/N $>$ 5.

With this in mind, we proceed with the cut [O{\small III}]$_{5007}$~S/N $>$ 3 for our complete [O{\small III} emitting sample to provide a large control group that spans the broadest range of stellar mass and star formation rate. However, when considering the incidence of outflows we instead use a cut of S/N $>$ 5 as this is the sample in which outflows are effectively detected.


As discussed, these criteria are not very strict and as a result our sample size reduces only slightly to 1087 galaxies at S/N $>$ 3 and 1007 at S/N $>$ 5.

\subsection{Consideration of NIRSpec Line Spread Function}\label{sec:mr-spec}

Before selecting our outflow candidates, it is imperative to consider the line spread function (LSF) of the medium resolution gratings and the broadening effect it can have on the emission lines in our sample. Several studies have worked to quantify the LSF across the varying resolutions and filters of the NIRSpec instrument (see \cite{2024A&A...684A..87D} for a comprehensive discussion of the high resolution gratings and \cite{2024arXiv240416963X} for the medium resolution gratings).

The LSF of individual objects depends on multiple factors including object size, object morphology and the position of the source in slit. Determining the true LSF for each object requires intensive modeling and lies outside the scope of this study. Therefore, we instead adopt the values determined in \cite{Isobe_2023} which are calculated by observing the broadening of the planetary nebula IRAS-05248-7007. They find an LSF that varies with filter and wavelength with values between $\sim$100-150km s$^{-1}$ for the filter/grating combinations used in this study.

Since these values are significantly broad, we ultimately consider a modified methodology to that of \cite{xu2023stellaragnfeedbackprobed} and \cite{2024A&A...685A..99C}. The complete description of this methodology is described below.

\subsection{Line fitting of medium resolution spectroscopy}\label{sec:lsf}

To build a collection of potential outflow hosts, we perform line fitting on the medium resolution data for each galaxy. Each object in our sample has medium resolution spectra in all three dispersers, but here we limit our initial analysis to the disperser containing the [O{\small III}]+H$\beta$ complex. In the case of an overlap between two filters, we calculate the median of the two spectra to retrieve an improved signal to noise. Before fitting, we mask emission lines and model the continua across the whole wavelength range with the \texttt{specutils} python package. We then subtract the modeled continua from each object, though we note that the majority of objects in our sample do not have any visible continua in the medium resolution spectroscopy so the overall subtraction is typically negligible.

With our continuum subtracted spectra, we first fit the [O{\small III}]$_{5007}$~line with a single Gaussian component. We set a strict prior on the center of the line according to the spectroscopic redshift determined from manually inspecting the appropriate medium resolution spectrum. We do this to account for the shift between the prism and medium resolution spectra noted in section \ref{sec:MSA}. We allow  amplitude and $\sigma$ to act as free parameters, but impose a maximum value on $\sigma$ of 200km s$^{-1}$. We do this to ensure our single Gaussian maps a narrow emission line where any broadening is produced by typical gas dispersion or the line spread function of the spectrum. We then repeat this process on the fainter [O{\small III}]$_{4959}$ line and the H$\beta$ line, restricting only the $\sigma$ to match that of the brighter [O{\small III}] line. In this way we can check for any systematic offsets between the three emission lines that may be further evidence for significant outflows. We ultimately do not find any such offsets and also do not find any significant residuals in the line fitting of [O{\small III}]$_{4959}$ that would be indicative of broadening. This is primarily due to the lower S/N of the [O{\small III}]$_{4959}$ where any broadening would likely fall below the noise level of the spectrum. 

To search for outflow candidates, we determine the residual of the narrow line model on the [O{\small III}]$_{5007}$~line. We then select any galaxy with a significant ($>$ 3$\sigma$) residual on the blue side of the center of the fit. We target the blue side of the line to ensure we are selecting asymmetries most likely to be caused by outflowing gas instead of general broadening through either the LSF or high dispersion in the gas velocity fields \citep{bicon-agn-1,bicon-agn-2,bicon-agn-3}. These steps result in the selection of 60 galaxies. While our approach aims to maximize the reliability of the outflow identification, we note that our strict criteria likely makes that some outflow candidates are missed.  Recent works have found that redshifted wings in the emission lines of high-redshift galaxies can also be associated with outflows \citep{uebler2023,2024A&A...685A..99C}, corresponding to cases where most of the outflow points away from the observer.

From our conservative sample we can select the strongest outflow candidates. To do this, we fit an additional Gaussian component (hereafter outflow component) to the initial narrow component to model each [O{\small III}]$_{5007}$~line. For this component, we introduce two priors. First that the central peak of the Gaussian sits on the blue side of the narrow component and second that the standard deviation of the Gaussian is broader than the narrow component by a factor $\geq$ 1.2. We introduce this second factor following Ca24 which is designed to decrease the likelihood of selecting rotating star forming clumps within the galaxy. Such clumps are likely to show emission line signatures similar to that of the main galaxy only blueshifted/redshifted based on the orientation of rotation. Thus they likely materialise as a second narrow peak which the 1.2 factor intends to exclude. This approach is not perfect however, and additional checks to our final sample are required (See Section \ref{sec:clumps}). We finally allow our other parameters to remain free to ensure the outflow component can model the line asymmetry appropriately without affecting the underlying narrow emission line.

\subsection{Selection of outflow candidates} \label{sec:out-v}

To select the strongest outflow candidates, we impose two final selection criteria. First we state that the introduction of the outflow component must improve the overall fit of the emission line. Quantitatively, we calculate the Bayesian Information Criterion (BIC) for both the single and double Gaussian fits as BIC = $\chi^{2}-k\cdot log(n)$ where k = 3 (6) is the number of degrees of freedom in the single (double) Gaussian model and n is the number of data points in the fit. We secondly compute the S/N of the fitted outflow component using the same approach stated in section \ref{sec:InitialSelect}. A galaxy then passes our selection if it meets the criteria:

\begin{itemize}
    \item $\Delta$BIC = BIC$_{Outflow}$ - BIC$_{Narrow}$ $>$ 10
    \item S/N$_{Outflow}$ $>$ 3
\end{itemize}

\begin{figure*}
    \centering
    \includegraphics[width=0.6\linewidth]{ 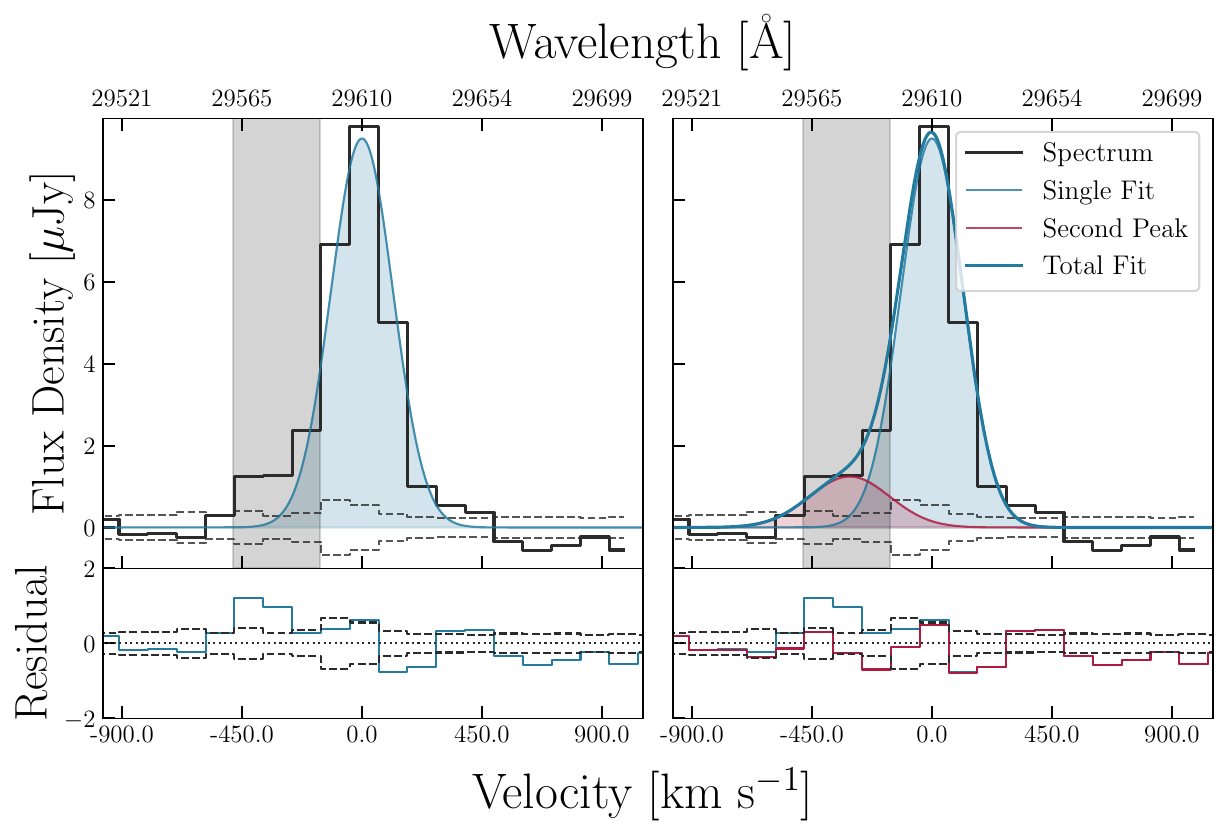}
    \caption{An example of our outflow candidate selection process. \textbf{Left:} The [O{\small III}]$_{5007}$ line of candidate JADES-GN+189.11339-62.22768 fitted with a single narrow Gaussian (blue). The flux excess on the blue side of the line is shaded grey. We see the flux (top panel) and residual (lower panel) are both significantly above the associated errors (grey dashed). \textbf{Right:} The same emission line fit with an additional broad component (red). This second component improves the residual whilst remaining significantly above the associated uncertainties.}
    \label{fig:fit-example}
\end{figure*}

This S/N threshold is in line with that of Ca24 and Xu23 and is decided as a compromise between the reduced depth of the medium resolution spectra and the desire to avoid identifying noise as outflow candidates. An example fit is shown in Fig. \ref{fig:fit-example} and applying these criteria reduces our sample to a total of 40 galaxies across the redshift range 2.5$<z<$9.1. We present the fitting parameters and fitted spectra for all our candidates in Appendix \ref{app:fit-tab} (see Tables \ref{tab:fit_params} and \ref{tab:fit_params_s}, as well as Fig. \ref{app:fit-all}).

We finally derive the outflow velocity from our fitting parameters using the relation defined in \cite{Rupke_2005} as

\begin{equation}
    v_{out} = \Delta v + 2\sigma_{outflow}
\end{equation}

where $v_{out}$ is the outflow velocity, $\Delta v = v_{outflow}-v_{narrow}$ is the peak separation of the two lines and $\sigma_{outflow}$ is the $\sigma$ of the outflow component. We further introduce a correction for the LSF such that $\sigma_{outflow}^{2}=\sigma_{broad}^{2}-\sigma_{LSF}^{2}$ where $\sigma_{broad}$ is the $\sigma$ of the broadened outflow component and $\sigma_{LSF}$ is the instrumental broadening. This value varies between 100-200km~s$^{-1}$ depending on the wavelength of the [O{\small III}] line \citep{Isobe_2023}.

\begin{figure}
    \centering
    \includegraphics[width=1.0\linewidth]{ 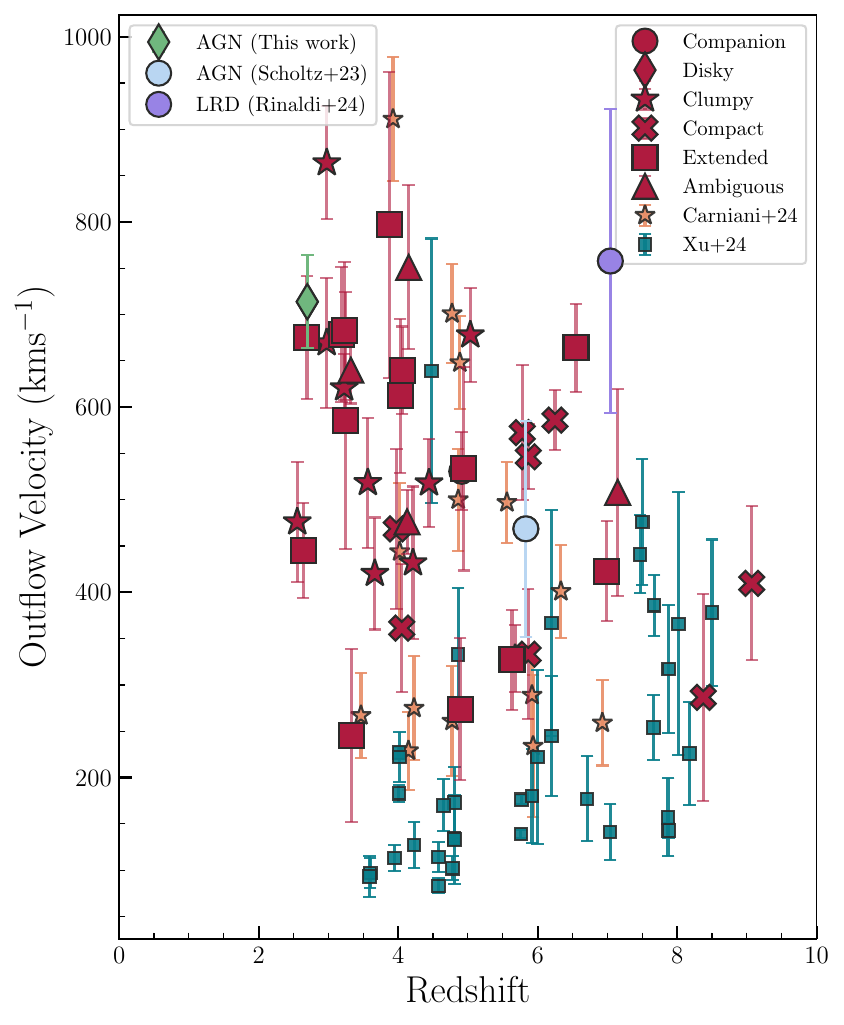}
    \caption{Outflow velocities derived from our line fitting procedure against spectroscopic redshift. The various marker shapes represent the morphology groupings described in the text. Included as orange stars and blue squares are outflow candidates from Ca24 and Xu23 respectively. The two AGN candidates and single LRD identified in this study, \citep{scholtz-agn} and \citep{rinaldi2024-lrds} are shown in green, blue and purple respectively.}
    \label{fig:outflow-z}
\end{figure}

The distribution of our calculated outflow velocities against redshift is shown in Fig. \ref{fig:outflow-z} where we also group objects by similar morphologies (see Section \ref{sec:morph}). We find a median outflow velocity of 531$^{+146}_{-159}$~km~s$^{-1}$, substantially higher than that of other studies across the same redshift range (\citealt{out-v-1}; Xu23; Ca24), although we note that we are limited to higher velocities by the medium resolution spectra. A broad anti-correlation between outflow velocity and redshift seems apparent, but we caution this relation has a large scatter and ultimately suffers from a lack of candidates at higher redshifts. When considering the full dynamic range of outflow velocity offered by higher resolution studies, the scatter of this correlation increases further and we ultimately consider there to be no relation between the two values.

For the overall incidence of outflows, \cite{2024A&A...685A..99C} identifies an increase in the number of outflows with S/N, with less than 25\% of objects having a S/N $< 20$ and approximately 50\% of their sample having a S/N $> 50$. For our sample, we find a much larger 40\% of objects with S/N $< 20$ and a far lower 7.5\% of objects with S/N $> 50$. These differences may arise due to several reasons including the fact that we start with a far larger initial sample as well as the difference in selection criteria and the range of outflow velocities probed by the medium resolution spectra. We also note that our initial S/N $> 3$ cut does not include any additional sources and increasing this cut to S/N $> 5$ would not remove any sources from our sample.

\subsection{Identification of rotation driven broadening}{\label{sec:clumps}}
Whilst every effort has been made to ensure our sample consists of exclusively outflow hosting candidates, broadening through strong rotations can introduce similar profiles to those identified in our selection process. These rotations can arise from star forming clumps within a galaxy or, in the case of more massive galaxies, strong rotations in a galaxy's center due to a particularly steep rotation curve like those seen in local spiral galaxies \citep{Sofue_2001}.

\begin{figure}
    \centering
    \includegraphics[width=1\linewidth]{ 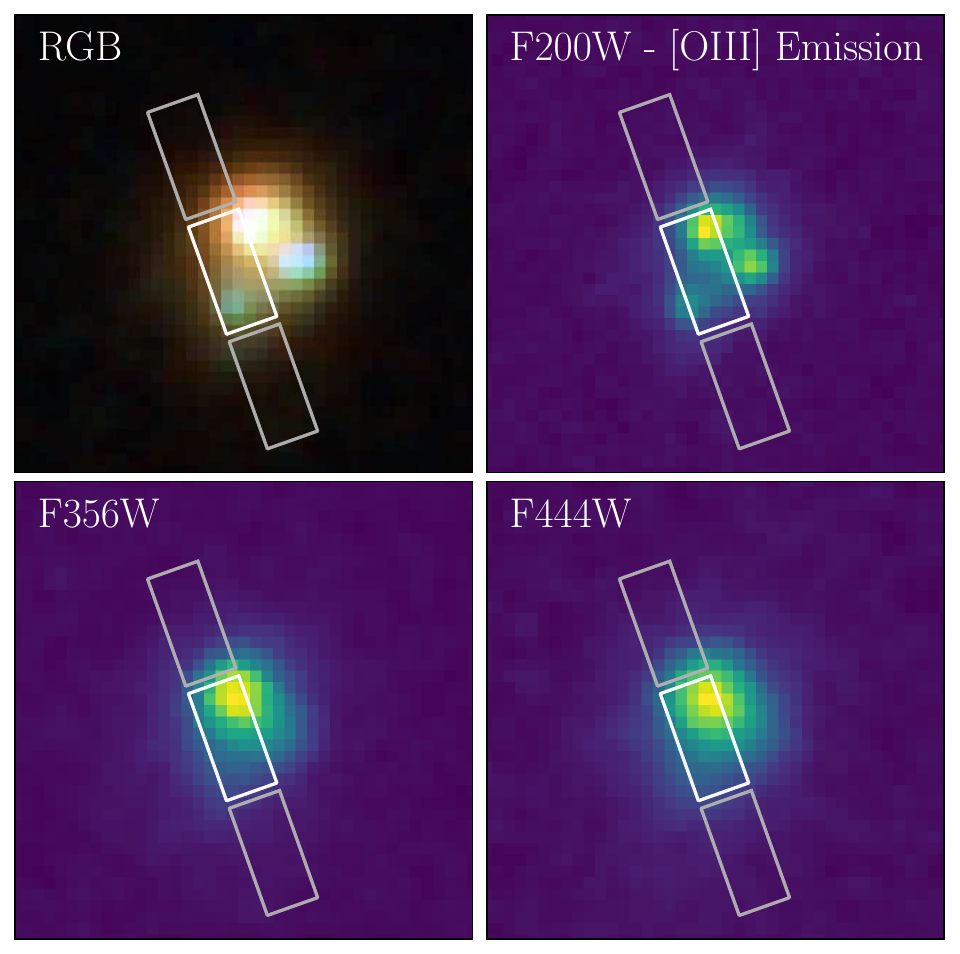}
    \caption{The NIRSpec slit orientation over-plotted on one of the objects in our sample. Multiple individual features are clear across the RGB image the F200W filter and NIRSpec slit overlaps all of them. Since the F200W filter covers the [O{\small III}] emission of this object, it is possible for the broadened emission line to be a result of these clumps.}
    \label{fig:clump} 
\end{figure}

As described in Section \ref{sec:lsf}, our fitting procedure is designed to exclude rotating clumps that seemingly broaden the host emission by introducing a secondary, shifted narrow component. However, with our selection criteria it is possible to still select low intensity, rapidly rotating clumps that produce similar broadened components to that of outflows. The simplest way of eliminating such objects is to visually inspect the NIRSpec slit orientation across each candidate. If the slit passes over two or more distinct clumps, we cannot rule out the possibility of a rotating clump leading to a false outflow detection. An example slit orientation is shown in Fig. \ref{fig:clump}. Here, the RGB image and F200W image present multiple clumps covered by the NIRSpec slit. Since the F200W filter contains the [O{\small III}] emission for this object, we cannot rule out the possibility that the broadened emission line is a consequence of multiple rotating clumps.

Among the objects with a clumpy morphology or those with nearby companions (see Section \ref{sec:morph}) 6 have NIRSpec slit orientations that either fully or partly overlap multiple clumps. Out of these 6 objects, none of them have detected companion objects. We note, however, that only 2 out of the 6 objects have line profiles preferentially modeled by two narrow peaks typically associated with rotating clumps, rather than a broadened outflow component.

Meanwhile for the consideration of rotating disks, we first note that such structures are suspected to be supported typically at stellar masses $\text{M}_{*} \gtrsim 10^{9} \, \text{M}_{\odot}$ \citep{2019MNRAS.490.3196P,10.1093/mnras/stac3712}, corresponding to $\sim$28\% of our sample. Recent high redshift ($z > 4$) results have found high velocity rotating disks ($v_{rot} \gtrsim 300$~km~s$^{-1})$ in significantly massive galaxies using [C{\small II}] \citep{filippo-rot-1, rowland2024rebels25discoverydynamicallycold} and H$\alpha$ emission \citep{2024ApJ...976L..27N}. Such velocity values may be detected in the medium resolution spectra, but we expect the footprint to be more biconical and are likely excluded from our sample as a result. For completeness, amongst the objects in our sample with stellar mass $\text{M}_{*} \gtrsim 10^{9} \, \text{M}_{\odot}$, we find that none has a significant redshifted component as a possible signature of rotation. Extending to the whole [O{\small III}] emitting sample, we identify 11 candidates above the same stellar mass cut that would pass a similar selection criteria if were we to include redshifted features.

\subsection{Identification of AGN candidates}{\label{sec:agn-rem}}

When identifying the sources of outflows within our sample, one must first account for possible AGN activity. Indeed, JWST has revealed an abundance of AGN at high redshift through photometry \citep{yang-agn,Kokorev_2024-agn}, NIRCam wide field slitless spectroscopy \citep{Kocevski_2023-agn} and NIRSpec observations \citep{Harikane_2023-agn, scholtz-agn}, with observations of AGN host galaxies reaching $\sim$10\% in the redshift range 4 $<$ z $<$ 6 \citep{maiolino_agn}. Such AGN can drive significant outflows, and whilst our selection criteria is biased against type-1 AGN, we consider our sample and evaluate the likelihood of an AGN being present on a source by source basis. 

First we perform a conscious check of the Balmer lines H${\alpha}$ and H${\beta}$. For type-1 AGN we expect these lines to have a (biconical) broad component originated in the AGN broad-line region, which should not be seen in [OIII]. In any case, we perform the same emission line fitting process as described above for both lines and check for any substantial (3$\sigma$) excesses. We find no evidence of such broadening in the Balmer lines within our outflow-host sample, so can rule out the presence of a broad-line AGN among our outflow hosts.

\begin{figure}
    \centering
    \includegraphics[width=1\linewidth]{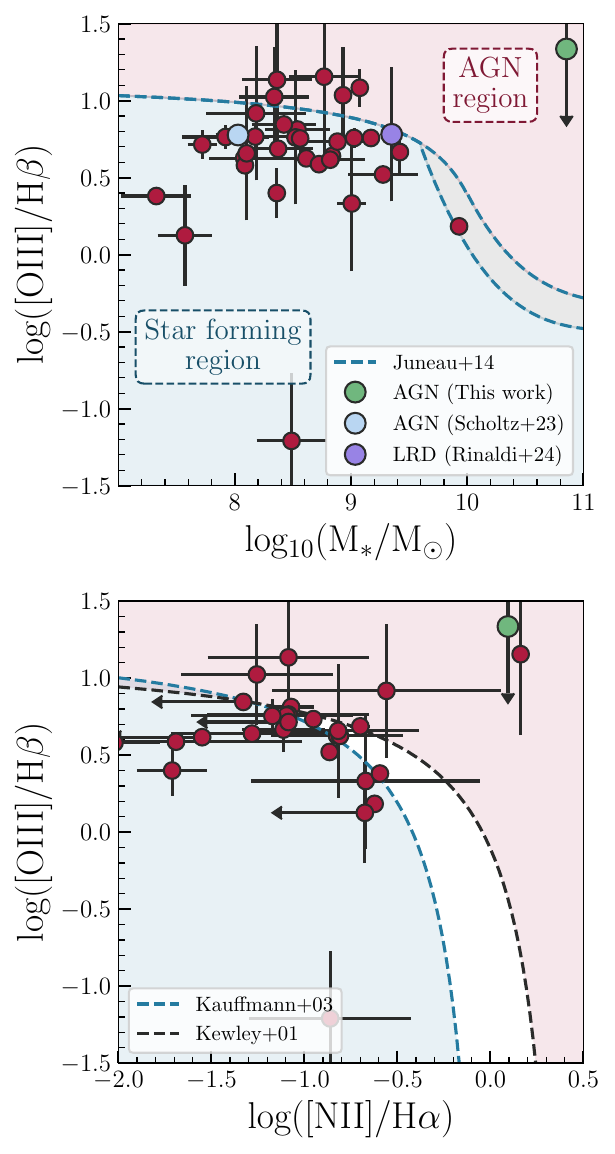}
    \caption{\textbf{Top:} The MEx diagram for the outflow candidates within our sample. Regions are taken from \cite{Juneau_2014}. The AGN region is shaded red whilst the star forming region is shaded blue. The composite region is shaded grey. \textbf{Bottom:}  The BPT diagram for the outflow candidates in our sample with detectable H$\alpha$. Separations are taken from \cite{kauffmann} and \cite{Kewley_2001}. AGN classification is the shaded red region whilst the star forming region is shaded blue. The composite AGN-SF region is left unshaded. Upper limits on line flux ratios are denoted by arrows. The only robust AGN candidate with high S/N emission lines is in green, whilst the AGN and LRD identified in \cite{scholtz-agn} and \cite{rinaldi2024-lrds} are blue and purple respectively.}
    \label{fig:bpt} 
\end{figure}

The process of identifying narrow-line, type-2 AGN is less straightforward. As a first check we make use of two diagnostic diagrams, the BPT diagram \citep{bpt-diagram} and MEx diagram \citep{mex-diagram}. To ensure the appropriate emission lines are individually resolved, we use values obtained from the fitting of the medium resolution spectra. We note however that in many cases the [N{\small II}] line is undetected and in a number of fewer cases the H$\beta$ line is also undetected. In these instances we provide upper limits extracted from the base level noise of the spectrum. The resulting line flux ratios are presented in Fig. \ref{fig:bpt}.

At first glance, it appears that several objects lie in the AGN region on the BPT diagram, though we caution the large errors on many of these candidates. We also note that, at high redshift, the BPT diagram struggles to classify AGN \citep{high-z-agn-ex1}. Particularly, star-forming galaxies at high redshifts tend to occupy the local AGN locus, while JWST has begun to uncover a population of high-redshift AGN that lie in the region occupied by low-redshift starburst galaxies \citep{maiolino_agn}. To some extent, the star-forming galaxy/AGN classification in the line-ratio diagrams is also blurred because of the effects of varying metallicities \citep{uebler2023} and, more generally, the evolution of interstellar medium properties with redshift \citep{backhaus2025}. Amongst the objects in our sample only one reliably shows the diagnostics associated with AGN whilst exhibiting high signal-to-noise emission lines.

Finally, we check the literature for known AGN in our sample. \cite{scholtz-agn} identified 42 galaxies in a sample of 209 from the GOODS-S JADES observations as type-2 AGN. Of these 42, a single object matches with our sample, JADES-GS+53.16685-27.80413. Rather than excluding this object, we distinguish it from the rest of our sample for the remainder of this work. For our compact sources, we additionally investigated the criteria devised to identify highly compact, red objects or the so-called ``Little Red Dots'' \citep[LRDs;][]{Matthee_2024-lrds}. These criteria involve a selection of colour cuts and compactness checks (See e.g., \citealt{labbe-lrd, Greene_2024-lrds, Kokorev_2024-agn, rinaldi2024-lrds} for details). Whilst our compact sources do indeed pass the compactness criteria, only one object passes the appropriate colour cuts to match the properties of observed LRDs. This object, JADES-GN+189.19835+62.29704,  is selected as an LRD in \cite{rinaldi2024-lrds}. Whilst not confirmed as an AGN, we also mark this object within our sample.

\section{Properties of outflow hosting galaxies} \label{sec:props}

Here we discuss the properties of our outflow hosting candidate sample derived from the combined photometric and spectroscopic data available. We place them in the context of the broader sample of [O{\small III}] line emitters and utilise imaging to identify the morphologies of our sample. We also describe the relations between outflow velocity and the various physical properties derived in this section.

\subsection{Outflow morphologies} \label{sec:morph}

\begin{figure*}
    \centering
    \includegraphics[width=1\linewidth]{ 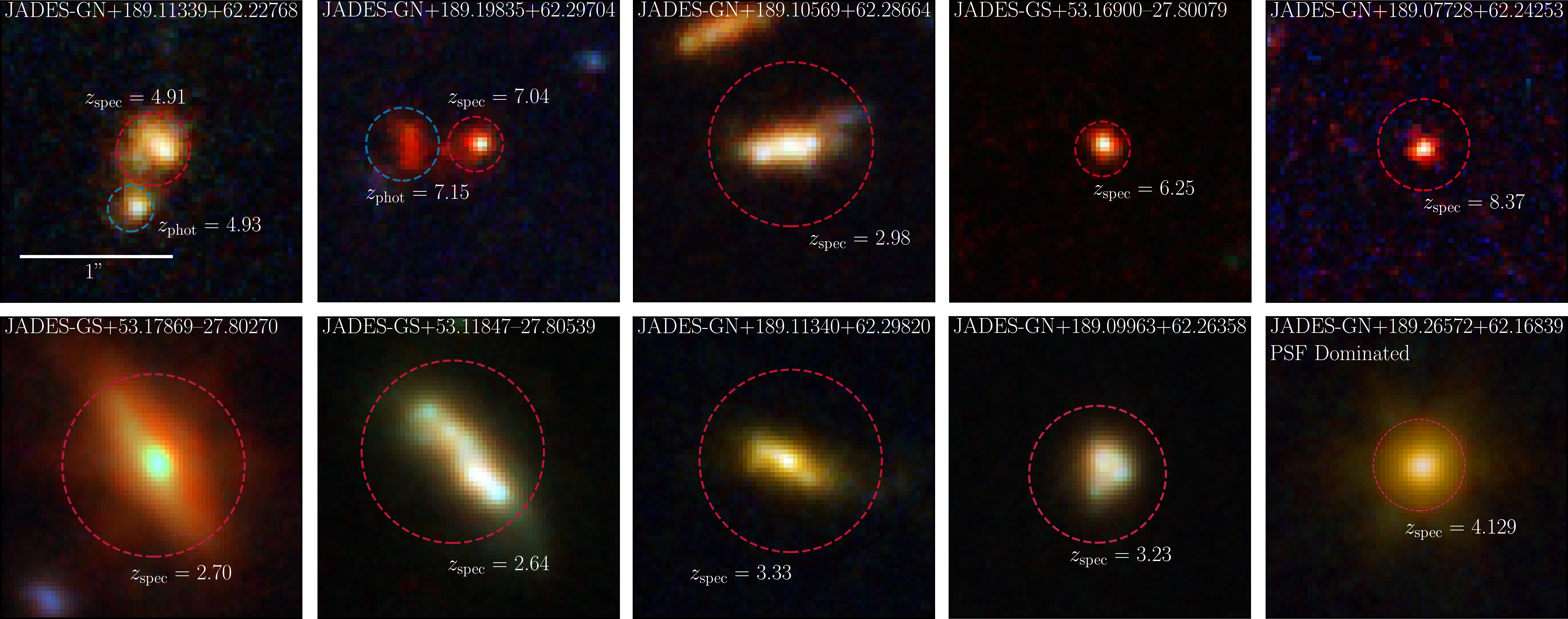}
    \caption{RGB image cutouts for a selection of outflow candidates with NIRCam coverage. Images are generated using all available photometric bands and the \texttt{trilogy} \citep{trilogy} package. Target galaxies are circled in red whilst nearby/companion objects are circled in blue. Spectroscopic redshifts are obtained from our line fitting whilst photometric redshifts come from the JADES public release photometric catalogs. We see a diverse range of morphologies ranging from clumpy or candidate mergers, extended or disc like galaxies or highly compact sources.}
    \label{fig:rgb-cut}
\end{figure*}

To better understand the sources of outflows in our sample, we study each object in all available NIRCam photometric bands. We build composite RGB images from the available filters to broadly characterise the morphology and environments of our outflow candidates. Fig. \ref{fig:rgb-cut} shows 2"  RGB cutouts for 10 of our sources across both the GOODS-N and GOODS-S fields selected to show the diverse range of morphologies exhibited.

In an effort to further constrain the outflowing gas, we perform a simple analysis to map the distribution of [O{\small III}] emission in our outflow candidates \citep{2024ApJ...970...19Z}. For each source, we identify appropriate filters that map the UV and optical continuum and one filter that best covers the wavelength of the [O{\small III}] emission. Optimally this would be performed with medium or narrow band imaging (see \citealt{medium-imaging-outflow} for a comprehensive study with medium band imaging), but the lack of coverage in both fields (limited to 4 medium bands in GOODS-N) restricts us to using wide band imaging in some cases. As such, whilst the emission line typically only boosts the flux in one filter, the surrounding continua is challenging to model. Nevertheless, we provide emission line maps by subtracting the filter associated with the optical continuum from the filter associated with the emission line. This allows us to  more stringently identify the morphologies of our sample whilst also tracing the location of the [O{\small III}] emission more directly. The number counts for each classification are shown in Table \ref{tab:morph_count} and Fig. \ref{fig:outflow-z} shows no apparent correlation between outflow velocity and morphology.

\begin{figure*}
    \centering
    \includegraphics[width=1\linewidth]{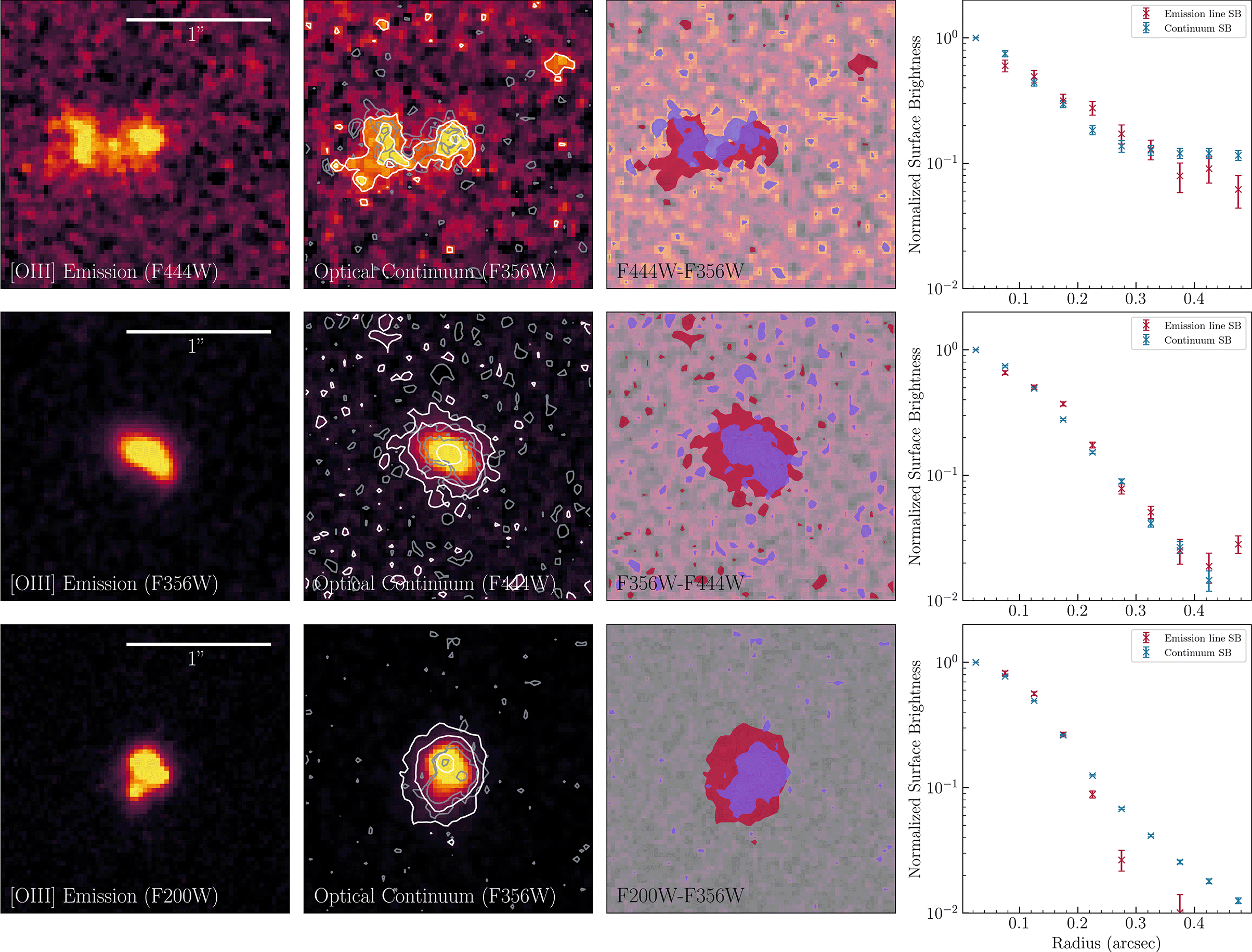}
    \caption{Example cutouts for three of our outflow candidates. \textbf{First column:} Cutout from the filter within which the [O{\small III}] emission line complex resides. \textbf{Second column:} Cutout from the nearest filter to the [O{\small III}] line that maps the optical continuum without including the line itself. White contours trace the optical continuum whilst grey contours trace the [O{\small III}] emission. \textbf{Third column:} Emission line map built from the subtraction of the optical continuum from the [O{\small III}] emission. The [O{\small III}] emission is blocked in blue whilst the surrounding continuum is blocked in red. \textbf{Fourth column:} The radial surface brightness profile of the continuum and emission line map. The surface brightness is normalised to the centre of each respective cutout.}
    \label{fig:em-map}
\end{figure*}

Simultaneously, we derive the surface brightness profile by averaging the flux in the image within 0.05" rings expanding from the center of each object. We normalise this flux to the central point of each source in the respective continuum and emission line images. From here we broadly categorise our sample based on the shape, environment, and surface brightness profile of each source. Example results are presented in Fig. \ref{fig:em-map} where we typically see clumpy or offset [O{\small III}] emission.

\begin{table}[t]
\centering
\caption{The distribution of objects within our outflow sample, including all clumpy sources where a distinction is made between those identified as rotating clump candidates. Ambiguous morphologies are those with low S/N imaging (typically in F444W only) or those dominated by the PSF.} \label{tab:morph_count}
\begin{tabular}{l|r}
\tablewidth{0pt}
\hline
\hline
\\
\textbf{Morphology} &  \textbf{Number Count}\\
\hline
Compact & 8 \\
Extended & 14 \\
Disky & 1\\ 
Clumpy (rotation candidates) & 10 (6)\\
Companion & 3 \\
Ambiguous & 4 \\
\hline
\end{tabular}
\end{table}

We see that the majority of our outflow candidates are extended or clumpy (e.g., 2nd row, cutouts 2-4 of Fig. \ref{fig:rgb-cut}), with 8 sources appearing as compact (e.g., Fig. \ref{fig:rgb-cut}: 1st row, cutouts 4 and 5) and three being dominated by the PSF e.g., Fig. \ref{fig:rgb-cut}: 2nd row, cutout 5). We also identify 3 objects with potential companions or secondary components with $\Delta z < 0.4$ for all cases (e.g., Fig. \ref{fig:rgb-cut}: 1st row, cutouts 1 and 2). None of the companion galaxies have spectroscopic coverage and their redshifts are determined from their multi-wavelength photometry. We differentiate between clumpy morphologies and sources with close companions by considering the segmentation images provided by the JADES team, where clumpy objects are identified as a single object and companions are distinct.

\subsection{SED Fitting}

To determine the physical properties of the objects in our sample, we perform SED fitting with the software \texttt{LePhare} \citep{lephare}. To provide a control sample to compare our outflow candidates with, we fit every object in our original [O{\small III}] emitting sample that has robust photometric coverage of at least 8 bands which populate the full filter range of NIRCam and HST.

For our SED fitting, we fix the redshift of every object in our initial sample to the redshift determined through our line fitting. We then follow a fitting procedure similar to that of \cite{navarrocarrera2024burstinesslowstellarmassha}. Namely, we utilise stellar templates from \cite{BC03} that incorporate an exponentially declining star formation history (SFH) whilst also allowing for bursty star formation in the form of a single stellar population (SSP). We then allow for young stellar ages (down to 10 Myr) and a solar (Z$_{\odot}$) or sub-solar (0.2Z$_{\odot}$) metallicity. We finally incorporate dust reddening from \cite{Calzetti_2000}, allowing E(B$-$V) values to vary from 0$-$1.5. An overview of our fitting parameters are presented in Table \ref{tab:fit_params}. We treat the clumpy sources identified in Section \ref{sec:morph} as whole objects whilst those identified with companions are considered separately.

For completeness, we also perform a combination of spectroscopic and photometric fitting with the \texttt{BAGPIPES} \citep{Carnall_2018} fitting code on our outflow candidates. \texttt{BAGPIPES} uses a Kroupa \citep{Kroupa} initial mass function (IMF), stellar templates from \cite{BC03} as well as additional emission lines from \texttt{Cloudy} models \citep{Cloudy, Cloudy2}. We utilise an exponentially declining SFH and fix the redshift of each object to the spectroscopic redshift determined in our line fitting. Priors on other parameters are left broad, allowing \texttt{BAGPIPES} to fully explore the parameter space for each galaxy.

Stellar masses from both SED fitting codes are in good agreement, with 78\% of objects within errors. We find however a larger scatter in star formation rates calculated from the UV magnitude (see Section \ref{sec:sfr}) and a lower fraction of starbursts (53\% vs 65\%). Whilst this discrepancy is non-negligible, we note that it does not ultimately affect the conclusions of this study.

\begin{table}[t]
\centering
\caption{\texttt{LePhare} fitting model and parameters for SED fitting of our [O{\small III}] emitters.} \label{tab:fit_params}
\begin{tabular}{l|r}
\tablewidth{0pt}
\hline
\hline
\\
\textbf{Parameter} &  \textbf{Reference/Prior}\\
\hline
Template & \texttt{BC03} \\
SFH & Exponential \& SSP \\
IMF & \cite{Chabrier} \\ \hline
Mass (log$_{10}$M/M$_{\odot}$) & 1$-$13 \\
Age (Gyr) & 0.001$-$13 \\
Metallicity (Z/Z$_{\odot}$)& 0.2;1 \\
Dust extinction & \cite{Calzetti_2000} \\
E(B$-$V) & 0.0$-$1.5 \\ 
\hline
\end{tabular}
\end{table}

\subsection{Physical parameters of outflow hosts}{\label{sec:sfr}}

To investigate the various physical parameters of our outflow candidates, we first estimate the star formation rate from each objects rest-UV ($\lambda_{rest}$ = 1500\AA) luminosity. We  determine the UV flux from photometry, using the measurement from the filter that encompasses the rest-frame 1500\AA \space emission. In instances where multiple filters cover this wavelength and these filters show a detection, the median is used.

\begin{figure*}
    \centering
    \includegraphics[width=1\linewidth]{ 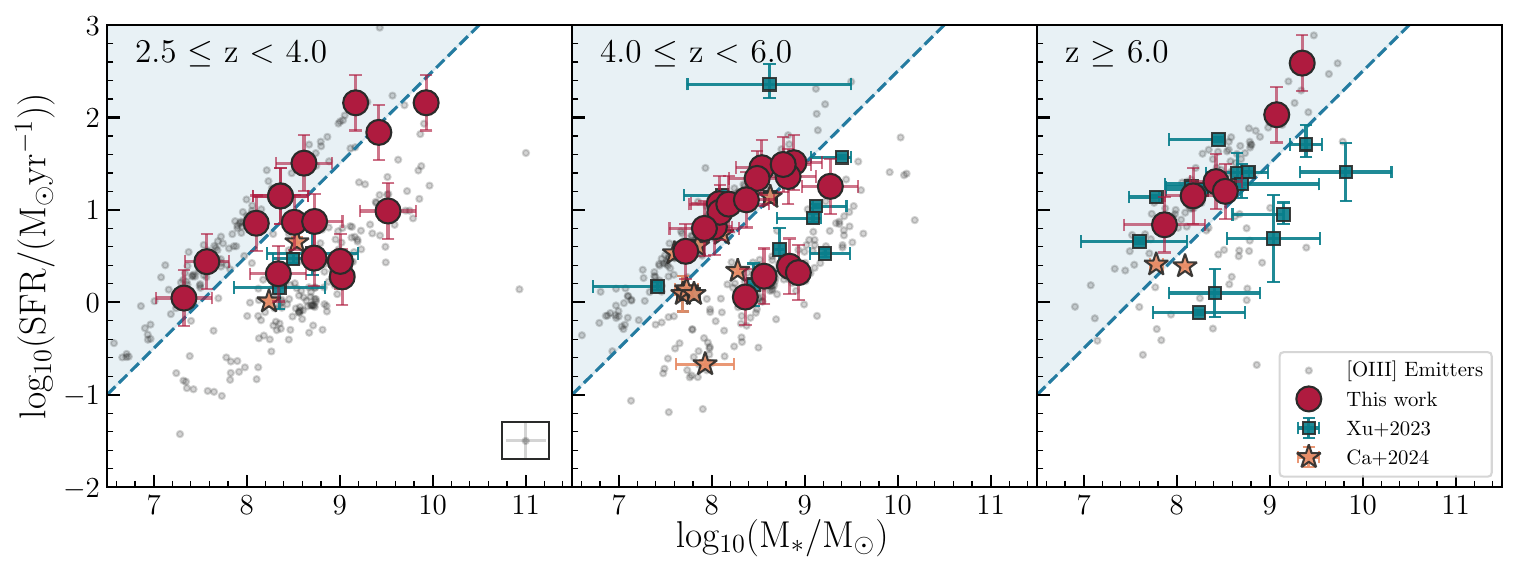}
    \caption{The SFR$-$M$_{*}$ plane for the [O{\small III}] emitters considered in this work binned by redshift. Candidate outflow galaxies are shown with the same red markers as Fig. \ref{fig:outflow-z} whilst the remaining [O{\small III}] emitting sample is shown as light grey points. Blue squares and orange stars are from Xu+24 and Ca+24 respectively. The lower envelope of the starburst regime as empirically defined by \cite{bimod1, Caputi_2021} is shaded in blue. Typical errors associated with our entire sample are boxed in the leftmost panel. The known AGN have been excluded from this figure as their SFRs are likely poorly estimated.} 
    \label{fig:mass-sfr}
\end{figure*}

We correct our UV fluxes following the dust-extinction law described in \cite{Calzetti_2000}, using E(B$-$V) values derived from our SED fitting. We then convert our fluxes to a monochromatic luminosity L$_{UV}$ and use the relation from \cite{Kennicutt_1998}

\begin{equation}
    \text{SFR}_{\text{UV}}(M_{\odot}yr^{-1}) = 1.4 \times 10^{-28} L_{UV} (\text{erg} \, \text{s}^{-1} \, \text{Hz}^{-1}).
\end{equation}

\noindent where we introduce a factor of 0.63 to account for our stellar masses following a Chabrier IMF \citep{Madau}. We show the calculated SFRs against stellar mass in Fig. \ref{fig:mass-sfr} where we probe a stellar mass range of $10^{6.5}< \text{M}_{*}/\text{M}_{\odot}< 10^{11.0}$ and  $0.1 < \rm SFR/\text{M}_{\odot}\text{yr}^{-1} \lesssim 100$ for our complete sample. Our outflow candidates exist in a lower dynamic range of $10^{7.5} \lesssim \text{M}_{*}/\text{M}_{\odot} < 10^{11.0}$ and $1.0 < \rm SFR/\text{M}_{\odot}\text{yr}^{-1} \lesssim 100$. We also find an average A$_{\text{V}}$ for our outflow sample of 1.60$^{+0.59}_{-1.6}$ and 1.18$^{+0.89}_{-1.18}$ for our complete [O{\small III}] emitting sample.

For the full [O{\small III}] emitter sample, we recover the bimodality between main sequence and starbursts similar to that seen in previous literature \citep{bimod1,bimod2,bimod3, bimod4}. The fraction of our outflow candidates classified as starbursts is found to increase with redshift and overall we find 68\% of our outflow candidates and 60\% of our full sample lay in the starburst regime. However, we caution that the total fraction of starbursts follows a similar trend and disentangling these two correlations is difficult without a significantly larger outflow sample. Despite this, we do note that at lower redshifts ($z\leq1.5$), recent results have found that strong outflows are more likely to reside away from the main sequence (e.g., \citealt{low-z-wind-1, low-z-wind-2,low-z-wind-3}; see \citealt{low-z-outflow-burst} for a complete summary).

\begin{figure*}
    \centering
    \includegraphics[width=1\linewidth]{ 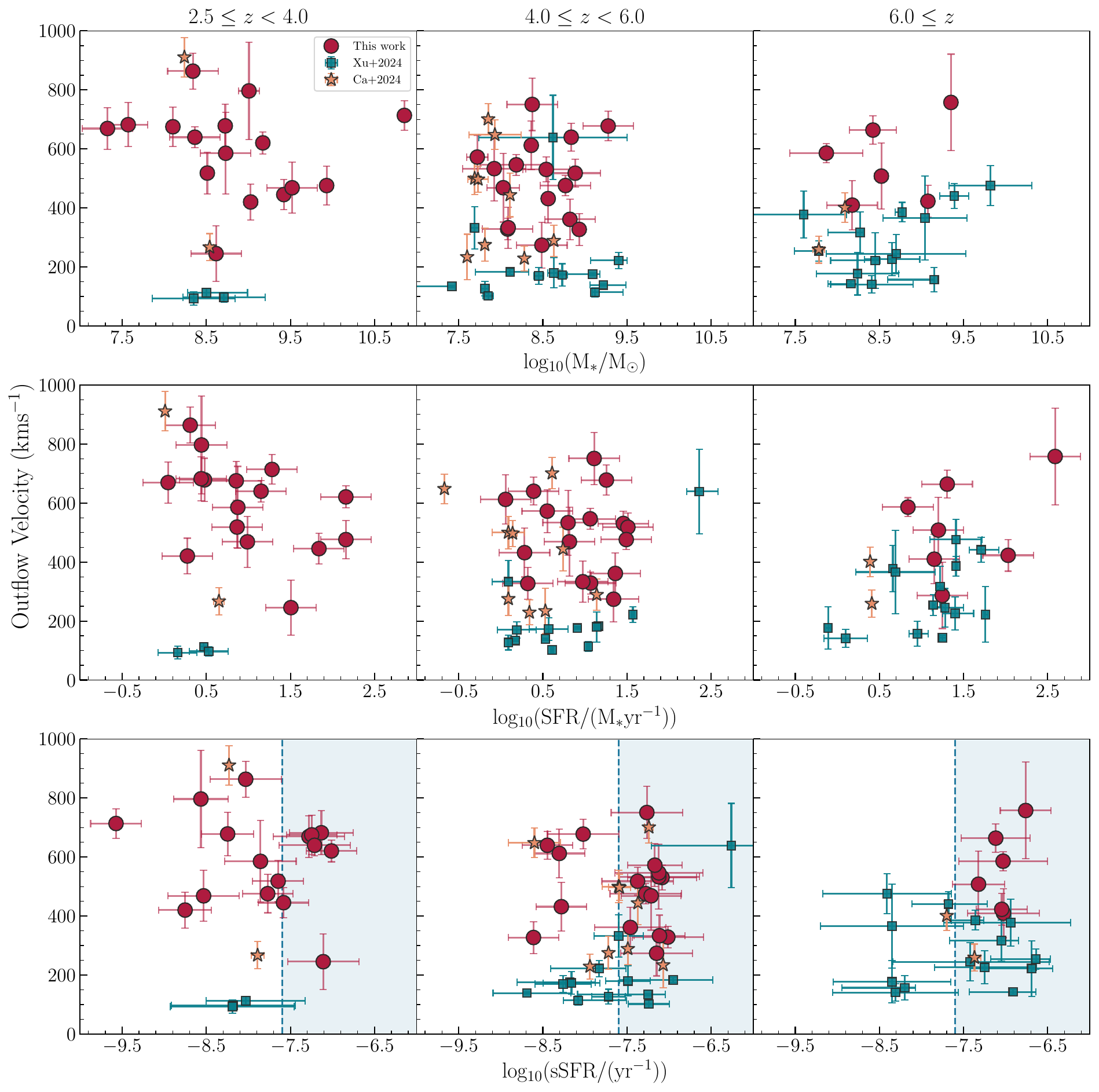}
    \caption{The derived outflow velocity for our sample against host physical parameters binned by redshift. \textbf{Upper:} Comparison with stellar mass; \textbf{Middle:} Comparison with UV star formation rate. \textbf{Lower:} Comparison with specific star formation rate derived from the UV SFR. Outflow candidates from Xu+24 and Ca+24 are shown in blue squares and orange stars respectively. The empirical starburst separation from \cite{bimod1,Caputi_2021} is shaded in blue.}
    \label{fig:out-v-phys}
\end{figure*}

We further compare the velocity of outflowing gas with the physical parameters of our outflow candidates using the outflow velocities derived in section \ref{sec:out-v}. These comparisons are shown in Fig. \ref{fig:out-v-phys} where we see no significant correlation between outflow velocity and stellar mass or SFR across any redshift bins. Previous studies at lower redshifts have found correlations with stellar mass (e.g., \citealt{low-z-mass-outv-1}) but such correlations are typically weak (e.g., \citealt{low-z-mass-outv, low-z-mass-outv-2}). In fact \cite{out-v-dyn-range} reports that, whilst these relations appear to exist in individual studies, compounding results to expand the dynamic range of stellar masses reveals that any relation has a significant scatter that indicates stellar mass is an unlikely contributor to outflow velocity. Our results are in good agreement with this statement as our stellar masses cover a considerable range of $10^{7.0} < M_{*}/M_{\odot} < 10^{10.0}$ (with one object at $M_{*}/M_{\odot} > 10^{10}$) and show no significant correlation.

This case for dynamic range is also prevalent for comparisons with SFR. Conversely, whilst we cover a significant range of masses, the range of SFRs covered by high redshift works is limited to predominantly $\log(\text{SFR}) > 0$. This is especially the case for spectroscopic results, where luminous emission lines associated with higher SFRs are favourably selected. Interestingly, there is an apparent positive trend between outflow velocity and SFR in the highest redshift bin (when also considering results from the literature), similar to that observed at lower redshift (e.g., \citealt{out-v-sfr-1,out-v-sfr-3,out-v-sfr-2}).  It is worth noting, however, that in this bin the object with highest SFR is the LRD reported by \cite{rinaldi2024-lrds}. This object greatly extends the overall range of SFRs in this bin allowing us to gleam at the potential underlying correlation, though care has to be taken this SFR is not being boosted by the presence of a black hole. Nevertheless, for other bins we lack the range of SFRs to identify any such trend. This phenomenon is again noted by \cite{out-v-dyn-range}. Extending this type of work to both higher and lower SFRs is therefore one of the key future aims in high redshift outflow studies.

Following \cite{out-sfrd}, we also briefly consider the relationship between outflow velocity and star formation rate density (SFRD) defined as 

\begin{equation}
    \Sigma_{\text{SFR}} = \frac{\text{SFR}}{2\pi r_e^2},
\end{equation}

\noindent where $r_e$ is the effective radius of the source obtained from the JADES public release data. However, we note that for our sources SFRD and SFR form a linear, positive correlation. As such any correlation (or lack thereof) with outflow velocity and star formation rate is mimicked by that of the star formation rate density.

Finally we importantly have to consider the dynamic range of outflow velocities in our study. Our work does not sufficiently probe low outflow velocities (v $<$ 200~km~s$^{-1}$) due to the limitations of medium resolution spectra. This, in combination with a low sample size when binned by redshift, may also obscure any meaningful correlations from being observed.

\section{Discussion} \label{sec:discussion}

We now consider the incidence of outflows within our sample and examine in detail selection effects, detection methods and limitations in the medium resolution spectra that effect the size of our sample. We also discuss the potential sources of these outflows, as well as comment on the consequences they may have on galaxy evolution. We do this by considering whether these outflows can escape their hosts and by determining the mass loading factor for our sources. Finally, we discuss some important caveats for this type of outflow modeling and compare our results to that of the literature.

\subsection{Outflow incidence and selection effects}\label{sec:seleff}

Given the nature of NIRSpec MSA observations having pre-selected targets, quantifying an outflow incidence fraction is challenging as accounting for this selection bias is practically unfeasible. Similarly, any initial selection criteria applied to a sample can distort the true fraction of outflows to non-outflows. For example, Xu23 identify an outflow fraction of 30/130 $\simeq$ 23\% but introduce an initial signal to noise cut of S/N $>$ 10. Similarly, Ca24 find an overall fraction of 25-40\% that varies with emission line strength. With our current selection criteria, we find a low outflow fraction of 34/1007 $\simeq$ 3.4\% (excluding rotating clump candidates) but increasing to S/N $>$ 10 improves the fraction to 32/476 $\simeq$ 6.7\% (where two objects are dropped due to the S/N). Our relatively low incidence fraction is similar to that observed in the local Universe. \cite{lowz-outflow, broad-comp-dust} report an outflow fraction of $\sim$7\% and $\sim$5\% respectively, when excluding AGN and considering outflow signatures in a similar velocity regime to our work. Our results are also lower than that observed at intermediate redshift. \cite{mid-z-incidence} find an overall fraction of 26\% at redshift 0.6$<$z$<$2.7, with a much more similar 11\% being contributed by stellar outflows. These studies notably make use of integrated field spectroscopy where gas emissions can be traced across the full spatial extent of each galaxy.

The primary drivers for our lower outflow incidence are the limitations of the NIRSpec medium resolution spectra and a stricter selection criteria.  As detailed in Section \ref{sec:mr-spec}, the medium resolution spectra may exhibit considerable emission line broadening from the line spread function. Since the overall effect of the LSF in combination with typical stellar velocity dispersions is difficult to quantify precisely, we adopted a broad lower estimate of 150~km~s$^{-1}$ with this value rising to $\sim$200~km~s$^{-1}$ at lower wavelengths. This alone prevents an accurate sampling of lower velocity outflows and prevents outflows of velocities $< 150$~km~s$^{-1}$ from being detectable entirely. 

Similarly, our stricter selection process results in a further decrease in sample size. Notably, we only select objects with asymmetry bluewards of the central emission line. We provide motivation for our selection criteria in section \ref{sec:out-v} but for completeness we repeat our initial selection allowing for  biconical broad components. In this way, we indeed find an initial increase in detections, but the broadened components do not surpass intrinsic broadening which can be caused by the LSF. Therefore we cannot comment with any confidence on their nature. As such, higher resolution spectra are needed to improve the accuracy in the derived overall fraction of outflows.

\subsection{Sources of outflows}

It is well documented that, whilst AGN are the origins of the strongest galactic outflows, star forming galaxies can equally produce galactic winds through stellar feedback \citep{Roberts_Borsani_2020}. As discussed in Section \ref{sec:sfr}, low redshift studies have described the increase in strength of outflows with increasing star formation. Whether this correlation remains true at higher redshift remains elusive due to sample size. As such it is important to consider all characteristics of our candidates in the hopes of providing some insight into the sources of their outflowing gas. Amongst the objects in our sample, two are classified as AGN and one is an LRD (see Section \ref{sec:agn-rem}). Two of these objects, the AGN identified by our line diagnostics and the LRD from \cite{rinaldi2024-lrds}, lie at least 3$\sigma$ above the typical outflow velocity of our sample, whilst the AGN identified in \cite{scholtz-agn} is within the expected range of our candidates. Worth noting is the fact that the LRD exhibits the fastest outflow of any object above $z=5$. 

We also consider that several of our outflow candidates are accompanied by companion objects typically of similar physical size and mass. In our sample, these paired systems are more abundant at higher redshift, all three at $z>$5. Our small sample size makes any comparison with observations \citep{low-z-mergers} and predictions \citep{merger-sim-EAGLE, merger-sim-horizon-agn, merger-sim-illustris} of close-pair fractions difficult and, whilst these objects may be merging events, there is the potential for highly irregular morphologies produced by bursty star formation \citep{burst-for-merger}. Past studies have posited that merging events at early times can promote galaxy wide outflows, either directly through merging \citep{merger-induced-outflows, merger-induced-outflows-2} or through the sudden growth of the central AGN (e.g., \citealt{merger-agn-growth}). 

In these cases we lack spectroscopic coverage to determine the nature of the companions. Thus, while the outflows in these systems may be enhanced by merging activity, trying to discern whether this is a direct impact of the merger or an effect from the growing AGN is difficult. This is especially important for the LRD in our sample which also has a close companion and may be undergoing a merging event \citep{rinaldi2024-lrds}.

\subsection{Consequences of outflows on galaxy evolution}\label{sec:cons}

\begin{figure*}
    \centering
    \includegraphics[width=0.7\linewidth]{ 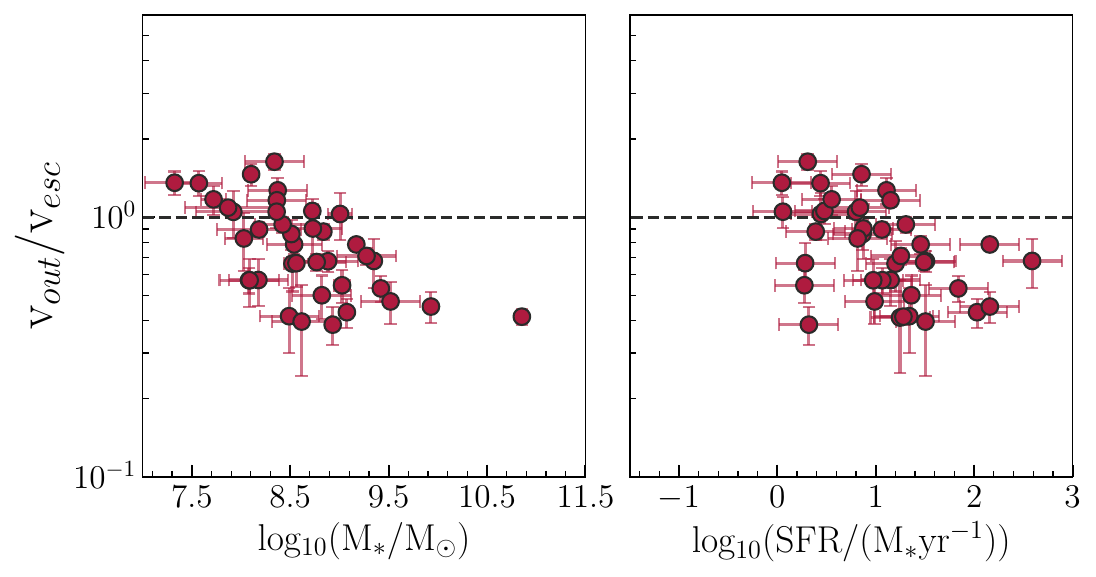}
    \caption{The ratio of outflow velocity vs escape velocity calculated assuming a spherical dark matter halo against stellar mass and star formation rate. The point of unity is plotted as a dashed line. Values above this value represent outflows that can escape the galaxy, whilst those below return to the interstellar medium. }
    \label{fig:vesc}
\end{figure*}

To gain an insight into the fate of the outflowing gas of our candidates, we determine the ratio of the outflow velocity to the escape velocity of each source. Recent studies have performed this calculation in different ways and here we adapt the method used in Xu23, though do not adopt the $v_{esc}=3v_{cir}$ approximation used in their study. Instead, we estimate the escape velocity ($v_{esc}$) of each source using the prescription from \cite{escape-v-eq}:
\begin{equation}
    v_{esc}=v_{200}\sqrt{2[1+\ln(r_{200}/r_{e})]},
\end{equation}
where $r_{200}$ is the radius within which the mean density is 200 times that of the critical density at the source redshift and $v_{200}$ is the circular velocity of the dark matter halo at this point. We retrieve the half light radius from the JADES public release catalogs and calculate the circular velocity and halo radius following (e.g., \citealt{2023ApJS..268...64W}):

\begin{equation}
    v_{200} = (\frac{GM_{h}}{r_{200}})^{1/2},
\end{equation}
\begin{equation}
    r_{200}=(\frac{GM_{h}}{100\Omega_{m}H_{0}^{2}})^{1/3} (1+z)^{-1},
\end{equation}
\noindent where we determine $M_{H}$ by following the $M_{*}$-$M_{H}$ relation detailed in \cite{m_star_mh_1} where we assume minimal evolution in redshift \citep{m_star_mh_2, m_star_mh_1}. A similar derivation is performed using the $M_{h}-M_{UV}$ presented in \cite{mh-muv} following Xu+3 (See Appendix \ref{app:muv}). We present the ratio of $v_{out}$ and $v_{esc}$ against stellar mass and star formation rate in Fig. \ref{fig:vesc}.

We see that, for the large majority of sources, the outflow velocity is smaller than the escape velocity, meaning outflowing gas is unlikely to escape the gravitational potential of our candidates. In this instance, the gas returns to the galaxy and we ultimately conclude that such outflows are ineffective at stripping the host of gas directly. Nonetheless these outflows are expected to have a direct impact on the galaxy interstellar medium. The derived velocities are still considerably high and the resulting turbulence likely heats and displaces the gas, leading to less efficient star formation (at least temporarily) \citep{wu2024ejectivefeedbackquenchingmechanism}. We further observe a negative correlation between the $v_{out} / v_{esc}$ ratio and stellar mass, but this is likely a consequence of the escape velocity scaling with halo mass and therefore also with stellar mass such that more massive hosts have greater escape velocities.

Our findings are similar to that of Xu23, but differ from the conclusions drawn by Ca24, who find a substantial fraction of sources with $v_{out} > v_{esc}$. This appears to be largely due to a different approach for calculating escape velocities. Xu23 compared the two approaches and finds the difference between these methods to be roughly a factor of 2. This factor would potentially change the conclusions made in this study and indeed a number of assumptions are employed regardless of methodology, so we caution that accurate modeling is required to truly understand the fate of this outflowing gas.

\begin{figure}
    \centering
    \includegraphics[width=1\linewidth]{ 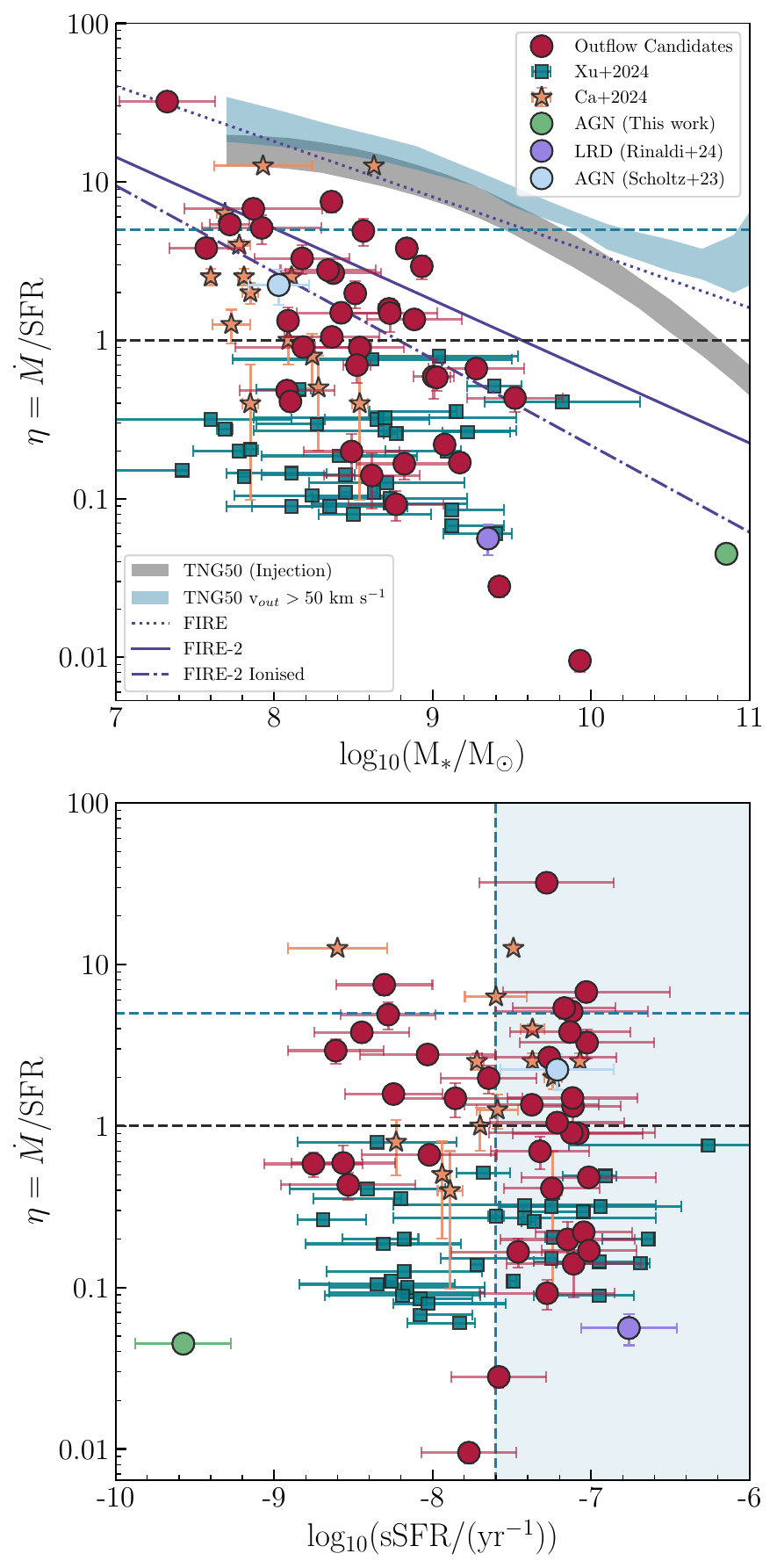}
    \caption{Mass loading factor against stellar mass (upper) and specific star formation rate (lower) for our outflow candidates. The horizontal line at $\eta=1$ represents the point at which the rate of ionising gas expelled from the galaxy is equal to the star formation rate. A horizontal line at  $\eta=5$ is also added for reference.  Results from Ca24 and Xu23 are shown as orange stars and blue squares respectively. The starburst regime defined by \citep{bimod1, Caputi_2021} is shaded in blue. Theoretical predictions from the TNG50 \citep{Nelson2019}, FIRE \citep{10.1093/mnras/stv2126} and FIRE-2 \citep{2021MNRAS.508.2979P} simulations are shown as shaded regions and purple lines respectively. }
    \label{fig:mass-load}
\end{figure}

Finally, we estimate the mass loading factor $\eta$ for our sample, defined as the ratio of mass-loss rate and star formation rate. This quantity determines if the amount of gas ejected from the host galaxy is more or less than that which is converted in the star formation process. A value of $\eta > 1$ suggests the gas outflow strips the galaxy at a faster rate than that of star formation, potentially producing a net quenching effect on the host. To determine our mass outflow rate, we adopt a similar approach to that of Ca24, with an outflow rate defined as:

\begin{equation}
    \dot{M}_{out} = M_{out}v_{out}/r_{out},
\end{equation}

\noindent where $v_{out}$ is the outflow velocity as defined in Section \ref{sec:out-v}, $r_{out}$ is the spatial extent of the outflow and $M_{out}$ is the mass of the outflowing gas (see e.g., \citealt{mass-outflow-eq} for a comprehensive definition). We follow Ca24 in setting $r_{out}$ to the radius of each source since measuring the true extent of the outflows requires detailed analysis and modeling of the 2D spectra. We then determine the outflow mass $M_{out}$ using the prescription from \cite{mass-outflow-eq-2}:

\begin{equation}
    M_{out}=0.8\times10^{8}(\frac{L^{corr}_{[OIII]}}{10^{44} \, erg \, s^{-1}})(\frac{Z_{out}}{Z_{\odot}})^{-1}(\frac{n_{out}}{500 \, cm^{-3}})^{-1} M_{\odot},
\end{equation}

\noindent where $L^{corr}_{[OIII]}$ is the dust corrected [O{\small III}] line luminosity, $Z_{out}$ and $n_{out}$ are the metallicity and electron density of the outflowing gas respectively. We follow Ca24 in setting $n_{out}=380~$cm$^{-3}$ and match the metallicity of the outflowing gas to that of the interstellar medium of each source determined through the empirical mass-metallicity relation defined in \cite{curti2023jadesinsightslowmassend}. We then determine the mass loading factor and plot the results against stellar mass and sSFR as shown in Fig. \ref{fig:mass-load}.

We derive a broad range of values for $\eta$, spanning almost four dex, and obtain a broad anti-correlation with stellar mass (upper panel of Fig. ~\ref{fig:mass-load}).   We note, however, that our single source with stellar mass $M_\ast > 10^{10} \, \rm M_\odot$ seems to suggest that the trend reverses at high stellar masses. As a matter of fact, such trend reversal is predicted by some galaxy formation models \citep{Nelson2019} and has also been derived from observational studies \citep[e.g., ][]{Concas2022}.  This corresponds to the change of regime where AGN outflows predominate (our single galaxy with stellar mass $M_\ast >10^{10} \, \rm M_\odot$  is indeed an AGN, according to the line ratio diagnostics; see Fig. ~\ref{fig:bpt}).  The other type-2 AGN in our sample has, instead, a low stellar mass and $\eta>1$ (note that for the computation of $\eta$ in these sources we assumed that the rest-frame UV luminosity could be converted into an SFR, i.e., it has negligible AGN contamination). 

For high stellar-mass galaxies our obtained mass loading factors are of the same order of magnitude as those derived by \citet[][see their Fig.~11]{Concas2022}. This is noteworthy, as their methodology to identify outflows is very different to that applied here (see Section~\ref{sec:caveats}). In addition, consistently with their finding, we find that the mass loading factors at high stellar masses are substantially lower than those predicted by galaxy formation models which consider ionised, neutral and molecular gas outflows altogether.  At low redshifts,   ionised gas has in general only a minor contribution, while the neutral and molecular gas phases typically dominate the outflow rate \citep{Fluetsch2021}. It is unclear whether this also holds at high redshifts.

 At stellar masses $M_\ast < 10^9 \, M_\odot$, our $\eta$ values become increasingly higher with decreasing stellar mass, but we note that the lack of low stellar-mass galaxies with low $\eta$ values could partly be a consequence of our selection effect preventing us to select outflows with $v_{out} < 200 \, \rm km/s$, so the observed anti-correlation must be taken with care. Our $\eta$ values  at low stellar masses are broadly consistent with the theoretical model predictions from FIRE-2 simulations that consider exclusively ionised outflows \citep{2021MNRAS.508.2979P}, though we note some scatter towards lower $\eta$ values.

At the same time, we observe a large scatter in the relation between $\eta$ and sSFR (lower panel of Fig.~\ref{fig:mass-load}). This scatter is likely a consequence of the two components making the specific star formation rate  (i.e., the stellar mass and SFR) sharing a negative correlation with the mass loading factor. As discussed in Section \ref{sec:sfr}, the star formation rate density of our sources closely traces the SFR, so we similarly report a decreasing $\eta$ with SFRD but acknowledge this is almost certainly driven by the definition of $\eta$. However, for completeness we note that some previous studies have found positive correlations with $\eta$ and SFRD (e.g. \citealt{int-outflow-4}), but similar studies and simulations have also reported the converse \citep{Li_2017,Kim_2020,eta-neg-rel}.

Galaxies with $\eta > 1$ are exhibiting outflows which expel ionised gas at a rate larger than the rate of star formation, thus potentially experiencing some sort of quenching effect. This quenching is unlikely to take the galaxy to a passive state, though, as the vast majority of known galaxies with stellar masses  $M_\ast< 10^9 \, \rm M_\odot$  at all redshifts are not passive and, when they are, the cessation of star formation is mostly driven by environmental effects.  Outflows are typically short-lived and their interaction with the galaxy interstellar and circumgalactic medium can be complex. It is much more likely that outflows are responsible for `mini-quenching', i.e. temporary quenching events, but the possible relation between the two should be tested via galaxy simulations. 

The dependence of $\eta$ on stellar mass has also been derived for low-redshift galaxies. For example, \citet{local-eta}
 found that the mass loading factor associated with ionised gas typically approaches unity for galaxies with stellar masses of $\sim 10^{11} \, \rm M_\odot$, while low stellar-mass starburst can reach much higher mass loading factors \citep[$\eta > 5$;][]{Heckmanetal2015}. For AGN, \citet{agn-eta} found that their typical mass loading factors are higher than those of starbursts at fixed stellar mass. In our sample, a few galaxies have $\eta \gtrsim 5$: they are all low stellar-mass galaxies, but none is an AGN.

\subsection{Possible caveats in the outflow identification via double Gaussian modelling}{\label{sec:caveats}}

The identification of gas outflows via a double Gaussian modelling of spectral lines and subsequent interpretation as it is performed here (and in a major part of the literature for high-redshift galaxies) implicitly assume that these gas flows originate at the centre of the galaxy and travel across a significant part of its interstellar medium. While this is likely a good approximation for small galaxies at high redshifts, the real outflow configuration can be much more complicated. In galaxy disks, outflows originate from star-forming regions located in different parts of the galactic plane and get out of the plane forming fountains, with the outflowing gas returning to the galactic plane in short timescales \citep[e.g., ][]{Fraternalietal2004, Fraternali2017}.

A rigorous extraction of the outflow parameters should be done by applying forward modelling with a number of free parameters, and extracting mock observations to be compared with real data which, in turn, allow for constraining the parameter values. This kind of approach has been adopted in some recent studies of outflows at low and intermediate redshifts \citep[e.g., ][]{Concas2022, Marasco2023}. These works derived mass loading factors which are substantially lower than the typical values reported in the previous literature, which they attribute to their different methodology for the outflow identification. 

When we compare with our own results, we find no obvious discrepancies with these previous works. As discussed above, the mass loading factors of our few highest stellar-mass galaxies are consistent with those of \citet{Concas2022}. For sources with $M_\ast \approx 10^9 \, M_\odot$, their upper limit on the mass loading factor is below our derived values, but it remains unclear whether this is due to the different methods to identify outflows, or a real effect mainly produced by the fact that our galaxy sample is at a significantly higher average redshift than the sample analyzed by \citet{Concas2022}. At the same time, the comparison with \citet{Marasco2023} must be done with care, as their mass loading factors refer only to the velocity component that exceeds the galaxy escape velocity. Future developments of forward modelling for the study of gas outflows in high redshift galaxies will clarify whether this methodology really leads to significantly different outflow derived parameter values.

\section{Summary and Conclusions}\label{sec:conc}

In this work, we utilised the complete sample of medium resolution NIRSpec MSA spectra from the JADES survey to identify and characterise high-velocity outflows in galaxies at redshifts $2.5 < z < 9$, from the presence of blue-sided broadening of their [O{\small III}]$_{5007}$~emission line.  Based on this spectroscopy and rich ancillary photometry in the GOODS fields, we find the following.

\begin{itemize}
    \item From the analysis and fitting of 1087 [O{\small III}]$_{5007}$~emission lines, we obtain
    a sample of 40 galaxies which exhibit a secondary blue-shifted peak relative to the central [O{\small III}] line. Out of these 40 galaxies, 34 are strong outflow host candidates whilst 6 sources have broadening potentially driven by rotating clumps.  Our robust outflow fraction of $34/1007 \simeq 3.4\%$ is considerably lower than other high redshift studies, which we account to the limitations of medium resolution spectra, and our stricter outflow selection criteria. When considering literature results from solely medium resolution spectra and increasing our S/N cut, our results become more inline with these other studies.
\end{itemize}

\begin{itemize}
    \item When we consider our final sample, we identify two type-2 AGN candidates, one from the line diagnostics performed in this work and one from the literature \citep{scholtz-agn}. A single LRD identified from the literature \citep{rinaldi2024-lrds} is also present in our sample. We also present 10 objects in our sample that show multiple components or clumps. Out of these, 6 candidates have NIRSpec slit orientations that overlap multiple clumps, but only 2 have line profiles preferentially modeled by two narrow peaks typically associated with rotating clumps.
\end{itemize}

\begin{itemize}
    \item The median outflow velocity of our sample is 531$^{+146}_{-159}$~km~s$^{-1}$. This is higher than previous studies but is artificially boosted by the typical broadening of the medium resolution spectra through the line spread function. This broadening limits the overall detectability of lower outflow velocities and makes any detection at v $< 150$~km~s$^{-1}$ impossible. 
\end{itemize}

\begin{itemize}
    \item No significant evolution between outflow velocity and redshift is present despite a declining median velocity of 584$^{+92}_{-61}$, 505$^{+42}_{-41}$, 531$^{+227}_{-46}$~km~s$^{-1}$ in redshift bins $2.5 < z \leq 4.0$, $4.0 < z \leq 6.0$ and $z > 6.0$ respectively. The high scatter of these values and the low number of high-z outflows (15\% of our candidates are at $z > 6.0$) makes any apparent correlation unreliable. Conversely, we see a marginal increase in outflow incidence from 2.4\% to 4.3\% and 4.4\% across the same redshift ranges.
\end{itemize}

\begin{itemize}
    \item We find that the majority of the outflows in our sample cannot escape the gravitational potential of their host galaxy (the median ratio between outflow velocity and escape velocity is $0.77^{+0.36}_{-0.32}$). Thus, these outflows are unlikely to enrich the surrounding circumgalactic medium by removing metals from the host galaxy. However, the generated turbulence, displacement of gas and heating of the interstellar medium will likely have an overall effect on the host galaxy. Larger statistical studies, as well as integral field spectroscopy,  at high redshifts are needed to inform simulations in order to fully understand how these outflows influence their host star formation activity.
\end{itemize}

\begin{itemize}
    \item The morphology of our outflow candidates is diverse, consisting of largely irregular, diffuse galaxies across the whole redshift range. Within our sample we also find a small number of compact sources, including a compact source with a tentative outflow component at z$\sim$9. We further identify 3 merger candidates, i.e., sources with close companions in projection, with all three these sources being at z $\gtrsim$ 5. Spectroscopic data for the companions is necessary to confirm the association with the outflowing sources.
\end{itemize}

\begin{itemize}
    \item Our outflow candidates populate different regimes of the $\text{M}_{*}-\text{SFR}$ plane, with 65\% being starburst galaxies. The fraction of starbursts increases towards higher redshifts. This is likely associated with a general increase in starburst at high redshift, as well as potential selection effects prioritising higher SFR galaxies.
\end{itemize}

\begin{itemize}
    \item We observe no strong correlations between outflow velocity and stellar mass or SFR, except at the highest redshifts in the case of star formation rate, where we see a tentative positive correlation. Unfortunately the limited statistics at $z \geq6$ prevents us to obtain a strong conclusion on this point.  We posit that the lack of correlation at lower redshifts could be due to the lack of range in star formation rate values. We also caution that our overall lower range of outflow velocities may prevent us from determining correlations across our sample.
\end{itemize}

\begin{itemize}
    \item The majority ($\sim60\%$) of outflows in our sample expel ionised gas at a rate larger than the rate of star formation ($\eta > 1$), with a few displaying  $\eta > 5$, all corresponding to low stellar-mass ($M<10^{8.5} \, \rm M_\odot$) galaxies. We see no trend in $\eta$ with redshift, with median values of 0.91$^{+3.17}_{-0.78}$, 2.41$^{+4.75}_{-1.88}$ and 0.79$^{+1.50}_{-0.61}$ for the redshift bins $2.5 < z \leq 4.0$, $4.0 < z \leq 6.0$ and $z > 6.0$ respectively.
\end{itemize}

This study details a sample of high-velocity galaxy outflows by tracing [O{\small III}] emission in JWST NIRSpec medium resolution spectroscopy. We probe a redshift range of 2.5$< z \lesssim$ 9.0 and a stellar mass range of $10^{7.5} \lesssim \text{M}_{*}/\text{M}_{\odot} < 10^{11.0}$. We show that a diverse range of galaxy morphologies can house these outflows and conclude that the outflows themselves can have a considerable effect on the evolution of their hosts primarily through the movement and heating of gas. Across our probed redshift range we see no evolution in properties including outflow velocity and mass loading factor, but further statistics are importantly needed to confirm these outcomes. Perhaps most critically, we demonstrate that JWST has vastly improved our capabilities to identify and characterise galaxy outflows, but we note that considerable work is still required to build a large statistical sample across the full dynamic range of redshift, stellar mass and star formation rate. Future wide field observations will require deep spectroscopy with medium or high resolution instruments and should be complemented or followed up by integrated field spectroscopy. This approach will allow for the most comprehensive analysis of outflows into the earliest epochs. \\

\section{Acknowledgments}

We thank Filippo Fraternali and Roberto Maiolino for insightful discussions. RAC, KIC, GD and RNC acknowledge funding from the Dutch Research Council (NWO) through the award of the Vici Grant VI.C.212.036. KIC and EI acknowledge funding from the Netherlands Research School for Astronomy (NOVA). This work is based on observations made with the NASA/ESA/CSA James Webb Space Telescope. The data were obtained from the Mikulski Archive for Space Telescopes at the Space Telescope Science Institute, which is operated by the Association of Universities for Research in Astronomy, Inc., under NASA contract NAS 5-03127 for \textit{JWST}. These observations are associated with \textit{JWST} programs GTO \#1180/\#1181, GO \#1210, GO \#1286 GO \#3215, GO \#1963 and GO \#1895. The authors acknowledge the team led by coPIs D. Eisenstein and N. Luetzgendorf, C. Williams and P. Oesch, for developing their respective observing programs with a zero-exclusive-access period.

\vspace{5mm}
\facilities{HST, JWST}

\software{astropy \citep{Astropy},
          BAGPIPES \citep{Carnall_2018},
          LePhare \citep{lephare}, 
          NumPy \citep{numpy},
          MSAEXP \citep{brammer_2023_8319596},
          SciPy \citep{Astropy},
          Topcat \citep{taylor2017topcatworkingdataworking},
          Trilogy \citep{trilogy}
          }

\appendix

\section{Escape velocity with M$_{UV}$}\label{app:muv}

\begin{figure*}[h]
    \centering
    \includegraphics[width=0.6\linewidth]{ 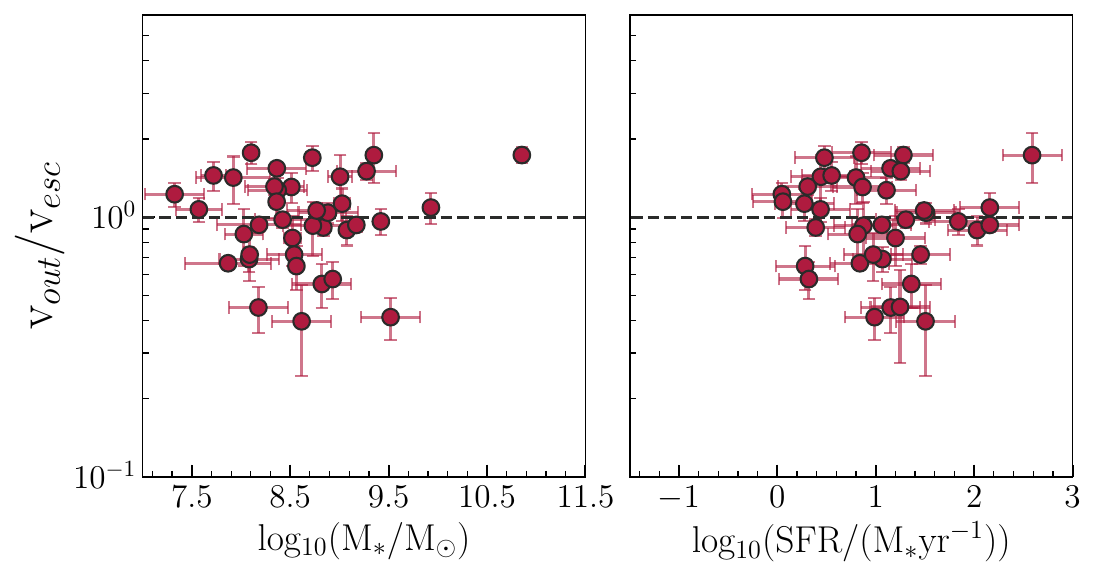}
    \caption{The ratio of outflow velocity and escape velocity against galaxy physical parameters. The escape velocity is calculated through an empirical relationship between UV magnitude and dark matter halo mass.}
    \label{fig:vout-muv}
\end{figure*}

Here we present an alternative figure for the ratio of outflow velocity to escape velocity, adopting the M$_{H}$-M$_{UV}$ presented in \cite{mh-muv} and used in Xu23. All other equations remain consistent with the main study (see Section \ref{sec:cons}). The results are shown in Fig. \ref{fig:vout-muv} where we see a slightly different result to that of  our main result. Namely, a number of objects now exhibit outflows that can escape their hosts. The relationship with stellar mass is no longer present since the derived halo masses no longer scale with stellar mass and instead scale with UV magnitude.

\section{[O{\small III}]$_{5007}$~emission line best Gaussian fit parameters and Gaussian fits}\label{app:fit-tab}
\newpage
\begin{table*}[h]
\movetabledown=2.5in
\begin{rotatetable*}
\begin{center}
\begin{tabular}{c|c|c|c|c|c|c|c|c|c}

\hline
\hline
\\
\textbf{Object Name}  & $z_{spec}$ & BIC & A$_{1}$ ($\mu$Jy)  & $\mu_{1}$ ($\mu$m) & $\sigma_{1}$ (\AA)  & A$_{2}$ ($\mu$Jy) & $\mu_{2}$ ($\mu$m)  & $\sigma_{2}$ (\AA)  & v$_{out}$ (km s$^{-1}$) \\

\hline
JADES-GN+189.15825+62.22136 & 9.070 & 86.766 & 6.161 $\pm$ 1.417 & 5.044 $\pm$ 0.908 & 16.829 $\pm$ 1.515 & 0.616 $\pm$ 0.099 & 5.042 $\pm$ 1.613 & 25.235 $\pm$ 7.066 & 409.651 $\pm$ 83.065 \\

JADES-GN+189.11339+62.22768 & 4.912 & 70.387 & 9.508 $\pm$ 0.951 & 2.961 $\pm$ 0.681 & 11.519 $\pm$ 1.728 & 1.246 $\pm$ 0.112 & 2.958 $\pm$ 0.385 & 14.874 $\pm$ 2.529 & 530.602 $\pm$ 42.112 \\

JADES-GN+189.14181+62.25841 & 4.941 & 25.137 & 7.204 $\pm$ 0.792 & 2.975 $\pm$ 0.268 & 10.942 $\pm$ 2.517 & 0.537 $\pm$ 0.145 & 2.974 $\pm$ 0.773 & 24.084 $\pm$ 2.408 & 533.291 $\pm$ 109.711 \\

JADES-GN+189.15742+62.27361 & 5.677 & 21.562 & 1.315 $\pm$ 0.289 & 3.345 $\pm$ 1.037 & 18.504 $\pm$ 4.071 & 0.292 $\pm$ 0.061 & 3.342 $\pm$ 0.401 & 14.577 $\pm$ 2.915 & 328.598 $\pm$ 36.504 \\

JADES-GN+189.12559+62.27485 & 3.872 & 23.930 & 1.300 $\pm$ 0.143 & 2.440 $\pm$ 0.415 & 11.261 $\pm$ 2.027 & 0.300 $\pm$ 0.063 & 2.437 $\pm$ 0.707 & 20.348 $\pm$ 5.697 & 796.753 $\pm$ 165.373 \\

JADES-GN+189.10832+62.29320 & 4.888 & 19.042 & 4.000 $\pm$ 0.680 & 2.979 $\pm$ 0.447 & 9.935 $\pm$ 3.179 & 0.560 $\pm$ 0.118 & 2.979 $\pm$ 0.834 & 17.275 $\pm$ 2.418 & 273.868 $\pm$ 76.718 \\

JADES-GN+189.11340+62.29820 & 3.330 & 22.244 & 6.086 $\pm$ 0.669 & 2.169 $\pm$ 0.564 & 9.485 $\pm$ 1.518 & 0.365 $\pm$ 0.058 & 2.167 $\pm$ 0.260 & 13.558 $\pm$ 4.339 & 245.328 $\pm$ 93.716 \\

JADES-GN+189.20530+62.25078 & 6.994 & 100.318 & 6.033 $\pm$ 0.483 & 4.003 $\pm$ 0.400 & 16.563 $\pm$ 2.816 & 0.507 $\pm$ 0.147 & 4.001 $\pm$ 0.320 & 24.249 $\pm$ 4.607 & 422.777 $\pm$ 53.960 \\

JADES-GN+189.19835+62.29704 & 7.043 & 88.941 & 2.582 $\pm$ 0.310 & 4.029 $\pm$ 1.209 & 20.938 $\pm$ 6.072 & 0.842 $\pm$ 0.143 & 4.025 $\pm$ 1.208 & 40.544 $\pm$ 3.649 & 757.709 $\pm$ 163.652 \\

JADES-GN+189.07728+62.24253 & 8.374 & 22.812 & 1.345 $\pm$ 0.444 & 4.695 $\pm$ 0.610 & 18.089 $\pm$ 3.075 & 0.384 $\pm$ 0.092 & 4.693 $\pm$ 1.361 & 28.581 $\pm$ 8.860 & 286.555 $\pm$ 111.893 \\

JADES-GN+189.23681+62.24825 & 3.190 & 25.903 & 3.597 $\pm$ 0.324 & 2.098 $\pm$ 0.650 & 7.002 $\pm$ 1.821 & 0.362 $\pm$ 0.036 & 2.095 $\pm$ 0.440 & 10.503 $\pm$ 3.256 & 678.325 $\pm$ 73.329 \\

JADES-GN+189.20745+62.26445 & 4.058 & 26.994 & 5.531 $\pm$ 1.327 & 2.533 $\pm$ 0.380 & 9.187 $\pm$ 2.940 & 0.199 $\pm$ 0.054 & 2.530 $\pm$ 0.405 & 12.671 $\pm$ 2.281 & 639.859 $\pm$ 47.104 \\

JADES-GN+189.02327+62.24565 & 4.152 & 27.881 & 8.766 $\pm$ 2.717 & 2.580 $\pm$ 0.542 & 10.758 $\pm$ 1.721 & 0.403 $\pm$ 0.048 & 2.576 $\pm$ 0.799 & 14.161 $\pm$ 3.399 & 750.920 $\pm$ 88.598 \\

JADES-GN+189.09963+62.26358 & 3.226 & 97.770 & 6.090 $\pm$ 0.913 & 2.117 $\pm$ 0.318 & 13.829 $\pm$ 3.734 & 0.957 $\pm$ 0.086 & 2.115 $\pm$ 0.211 & 10.592 $\pm$ 1.695 & 620.618 $\pm$ 36.902 \\

JADES-GN+189.22445+62.23836 & 2.975 & 26.300 & 4.429 $\pm$ 0.620 & 1.991 $\pm$ 0.358 & 10.454 $\pm$ 1.777 & 0.187 $\pm$ 0.043 & 1.989 $\pm$ 0.517 & 11.063 $\pm$ 1.770 & 669.337 $\pm$ 70.261 \\

JADES-GN+189.10569+62.28664 & 2.977 & 45.407 & 1.774 $\pm$ 0.195 & 1.992 $\pm$ 0.179 & 9.052 $\pm$ 2.897 & 0.814 $\pm$ 0.065 & 1.987 $\pm$ 0.457 & 9.963 $\pm$ 2.092 & 863.739 $\pm$ 60.985 \\

JADES-GN+189.15941+62.29740 & 3.231 & 29.524 & 4.164 $\pm$ 0.999 & 2.119 $\pm$ 0.297 & 7.995 $\pm$ 0.640 & 0.296 $\pm$ 0.083 & 2.116 $\pm$ 0.571 & 11.776 $\pm$ 2.120 & 682.353 $\pm$ 74.707 \\

JADES-GN+189.20825+62.17810 & 5.781 & 30.928 & 1.916 $\pm$ 0.268 & 3.396 $\pm$ 0.713 & 13.292 $\pm$ 3.589 & 0.327 $\pm$ 0.033 & 3.392 $\pm$ 1.085 & 16.994 $\pm$ 3.229 & 572.452 $\pm$ 72.810 \\

JADES-GN+189.23510+62.20376 & 4.213 & 32.128 & 5.669 $\pm$ 1.871 & 2.611 $\pm$ 0.365 & 8.711 $\pm$ 0.784 & 0.343 $\pm$ 0.055 & 2.609 $\pm$ 0.365 & 14.518 $\pm$ 4.791 & 431.741 $\pm$ 82.512 \\

JADES-GN+189.22884+62.20400 & 6.550 & 33.049 & 4.374 $\pm$ 1.137 & 3.778 $\pm$ 0.945 & 17.221 $\pm$ 3.789 & 0.284 $\pm$ 0.028 & 3.773 $\pm$ 0.604 & 21.011 $\pm$ 2.731 & 664.040 $\pm$ 47.508 \\

JADES-GN+189.27570+62.16170 & 5.035 & 30.588 & 7.072 $\pm$ 1.414 & 3.023 $\pm$ 0.242 & 10.083 $\pm$ 1.815 & 0.254 $\pm$ 0.033 & 3.019 $\pm$ 0.453 & 16.805 $\pm$ 2.689 & 677.830 $\pm$ 50.527 \\

JADES-GN+189.26572+62.16839 & 4.129 & 66.564 & 1.304 $\pm$ 0.157 & 2.570 $\pm$ 0.668 & 12.126 $\pm$ 1.576 & 0.865 $\pm$ 0.199 & 2.567 $\pm$ 0.488 & 8.580 $\pm$ 2.488 & 476.135 $\pm$ 34.518 \\
\hline
\hline
\end{tabular}
\caption{Fitting parameters and associated errors for the [OIII]5007 line of our final outflow candidates in GOODS-N. All parameters are left free except the centre point of each emission line which is fixed based on the redshift of each source. A prior is placed the $\sigma$ of each line to ensure the fitted components are narrow, with any broadening being limited to typical stellar dispersion and effects from the line spread function. } \label{tab:fit_params}
\end{center}
\end{rotatetable*}
\end{table*}

\begin{table*}
\movetabledown=2.5in
\begin{rotatetable*}
\begin{center}
\begin{tabular}{c|c|c|c|c|c|c|c|c|c}

\hline
\hline
\\
\textbf{Object Name}  & $z_{spec}$ & BIC & A$_{1}$ ($\mu$Jy)  & $\mu_{1}$ ($\mu$m) & $\sigma_{1}$ (\AA)  & A$_{2}$ ($\mu$Jy) & $\mu_{2}$ ($\mu$m)  & $\sigma_{2}$ (\AA)  & v$_{out}$ (km s$^{-1}$) \\
\hline
JADES-GS+53.17869-27.80270 & 2.698 & 42.811 & 2.098 $\pm$ 0.567 & 1.852 $\pm$ 0.167 & 15.439 $\pm$ 4.014 & 1.454 $\pm$ 0.436 & 1.849 $\pm$ 0.240 & 9.265 $\pm$ 2.038 & 713.724 $\pm$ 50.015 \\

JADES-GS+53.16900-27.80079 & 6.249 & 41.789 & 5.448 $\pm$ 0.599 & 3.628 $\pm$ 1.016 & 12.102 $\pm$ 1.210 & 0.564 $\pm$ 0.147 & 3.624 $\pm$ 0.471 & 18.153 $\pm$ 1.815 & 585.978 $\pm$ 32.084 \\

JADES-GS+53.17750-27.80252 & 5.864 & 54.563 & 2.574 $\pm$ 0.643 & 3.437 $\pm$ 0.722 & 10.873 $\pm$ 3.045 & 0.320 $\pm$ 0.083 & 3.436 $\pm$ 1.099 & 17.203 $\pm$ 2.752 & 333.207 $\pm$ 69.929 \\

JADES-GS+53.14676-27.78772 & 3.566 & 83.343 & 10.500 $\pm$ 1.050 & 2.284 $\pm$ 0.480 & 9.061 $\pm$ 2.537 & 0.750 $\pm$ 0.188 & 2.281 $\pm$ 0.479 & 11.426 $\pm$ 3.314 & 518.362 $\pm$ 70.032 \\

JADES-GS+53.11847-27.80539 & 2.638 & 77.374 & 4.621 $\pm$ 1.525 & 1.819 $\pm$ 0.436 & 6.604 $\pm$ 0.991 & 2.126 $\pm$ 0.383 & 1.817 $\pm$ 0.418 & 9.390 $\pm$ 0.939 & 445.355 $\pm$ 51.503 \\

JADES-GS+53.13259-27.75240 & 3.974 & 322.275 & 5.500 $\pm$ 0.440 & 2.490 $\pm$ 0.373 & 8.306 $\pm$ 2.326 & 0.904 $\pm$ 0.208 & 2.488 $\pm$ 0.647 & 15.497 $\pm$ 2.789 & 468.470 $\pm$ 86.165 \\

JADES-GS+53.16061-27.77625 & 2.554 & 35.503 & 8.251 $\pm$ 1.238 & 1.775 $\pm$ 0.355 & 6.555 $\pm$ 0.918 & 0.727 $\pm$ 0.109 & 1.773 $\pm$ 0.213 & 8.878 $\pm$ 2.752 & 475.977 $\pm$ 64.975 \\

JADES-GS+53.09574-27.77481 & 4.442 & 417.305 & 12.863 $\pm$ 3.344 & 2.724 $\pm$ 0.872 & 17.033 $\pm$ 1.703 & 5.523 $\pm$ 0.994 & 2.721 $\pm$ 0.517 & 13.626 $\pm$ 2.044 & 518.004 $\pm$ 47.328 \\

JADES-GS+53.16610-27.82751 & 7.147 & 44.362 & 2.195 $\pm$ 0.329 & 4.081 $\pm$ 0.653 & 19.998 $\pm$ 2.800 & 0.635 $\pm$ 0.146 & 4.078 $\pm$ 1.346 & 23.961 $\pm$ 7.188 & 507.951 $\pm$ 111.478 \\

JADES-GS+53.17704-27.81382 & 2.693 & 63.939 & 1.535 $\pm$ 0.230 & 1.850 $\pm$ 0.259 & 8.865 $\pm$ 2.393 & 0.528 $\pm$ 0.174 & 1.849 $\pm$ 0.259 & 15.428 $\pm$ 1.234 & 675.230 $\pm$ 66.120 \\

JADES-GS+53.09282-27.78722 & 4.053 & 102.968 & 4.822 $\pm$ 0.964 & 2.531 $\pm$ 0.683 & 8.445 $\pm$ 1.436 & 1.688 $\pm$ 0.456 & 2.530 $\pm$ 0.379 & 12.664 $\pm$ 4.052 & 361.294 $\pm$ 69.060 \\

JADES-GS+53.18393-27.80759 & 5.866 & 128.390 & 1.535 $\pm$ 0.460 & 3.441 $\pm$ 0.378 & 17.215 $\pm$ 5.337 & 1.437 $\pm$ 0.172 & 3.438 $\pm$ 0.344 & 17.215 $\pm$ 2.582 & 546.344 $\pm$ 35.243 \\

JADES-GS+53.13383-27.75115 & 3.249 & 48.229 & 3.838 $\pm$ 1.152 & 2.128 $\pm$ 0.489 & 8.132 $\pm$ 1.545 & 0.789 $\pm$ 0.063 & 2.126 $\pm$ 0.595 & 17.742 $\pm$ 3.371 & 585.431 $\pm$ 138.700 \\

JADES-GS+53.19575-27.78278 & 3.322 & 20.699 & 5.641 $\pm$ 0.959 & 2.165 $\pm$ 0.714 & 12.323 $\pm$ 1.109 & 0.500 $\pm$ 0.055 & 2.162 $\pm$ 0.281 & 10.835 $\pm$ 1.409 & 639.819 $\pm$ 35.997 \\

JADES-GS+53.15613-27.82130 & 4.028 & 25.526 & 3.412 $\pm$ 0.887 & 2.518 $\pm$ 0.504 & 8.401 $\pm$ 1.428 & 0.314 $\pm$ 0.035 & 2.515 $\pm$ 0.754 & 12.601 $\pm$ 3.780 & 612.257 $\pm$ 83.012 \\

JADES-GS+53.16685-27.80413 & 5.831 & 1038.648 & 1.596 $\pm$ 0.192 & 3.421 $\pm$ 0.684 & 14.164 $\pm$ 3.683 & 0.489 $\pm$ 0.044 & 3.419 $\pm$ 1.060 & 25.764 $\pm$ 2.576 & 468.575 $\pm$ 116.348 \\

JADES-GS+53.18817-27.80149 & 5.633 & 27.502 & 2.448 $\pm$ 0.294 & 3.322 $\pm$ 0.299 & 14.242 $\pm$ 1.567 & 0.816 $\pm$ 0.114 & 3.321 $\pm$ 0.432 & 19.160 $\pm$ 3.449 & 327.306 $\pm$ 54.001 \\

JADES-GS+53.18854-27.77870 & 3.666 & 22.770 & 1.489 $\pm$ 0.253 & 2.337 $\pm$ 0.514 & 7.796 $\pm$ 2.339 & 0.300 $\pm$ 0.030 & 2.335 $\pm$ 0.280 & 11.808 $\pm$ 3.306 & 420.297 $\pm$ 60.471 \\

\hline
\hline
\end{tabular}
\caption{Continuation of Table \ref{tab:fit_params} for candidates in GOODS-S.} \label{tab:fit_params_s}
\end{center}
\end{rotatetable*}
\end{table*}

\begin{figure}
    \centering
    \includegraphics[width=1\linewidth]{  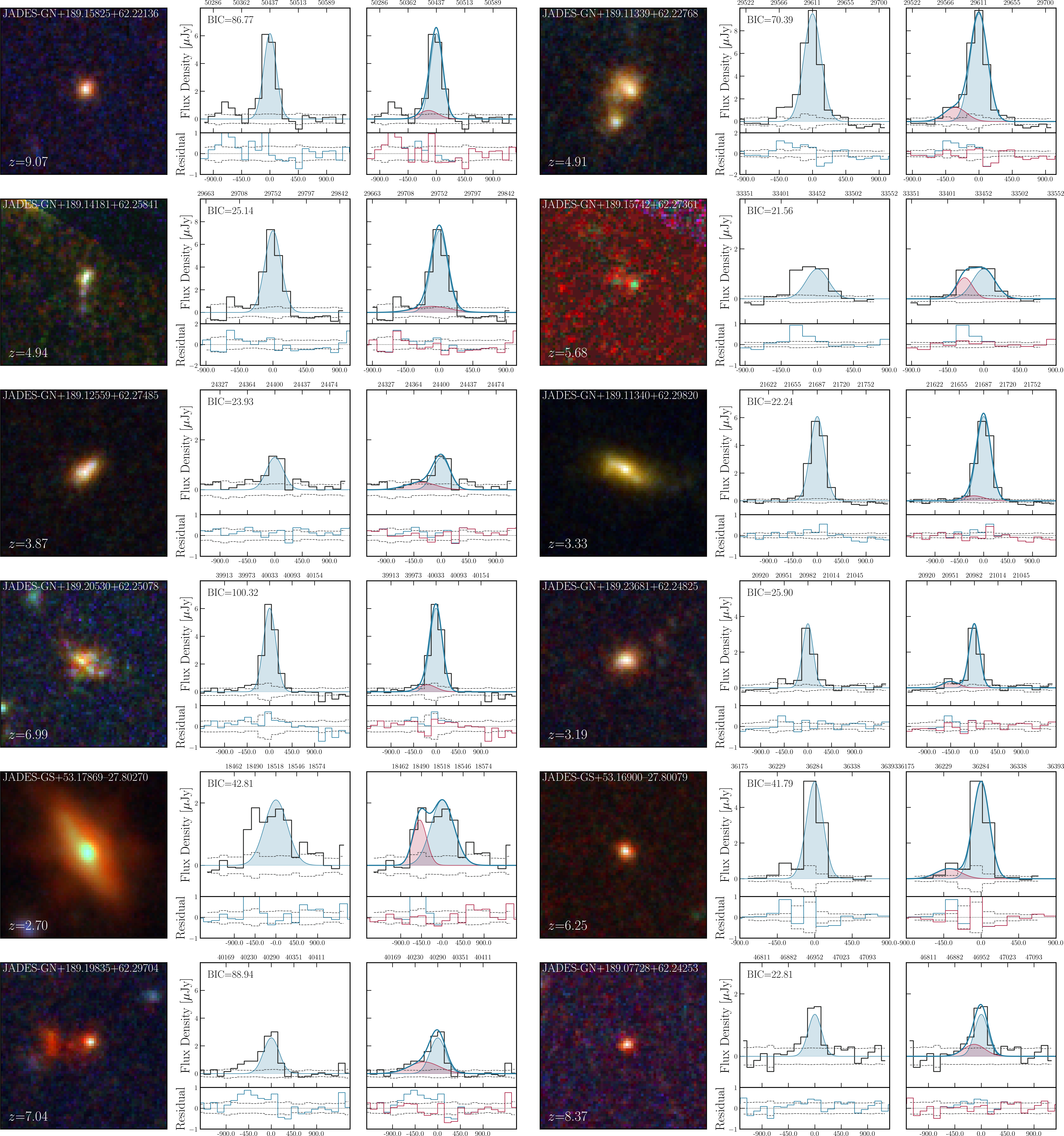}
    \caption{Gaussian fits of all outflow candidates in our sample including those produced by rotating clumps.\textbf{Left: } 1$^{\prime\prime}$ RGB postage stamp of the outflow candidate. These stamps are created following the same method as described in Fig. \ref{fig:rgb-cut}. \textbf{Center: }The [OIII]5007 line fitted with a single narrow Gaussian (blue). \textbf{Right: }The same emission line fitted with an additional broad component (red). Residuals for both fits are seen in the lower panels in blue and red respectively. \\ \\ \\ \\ }
    \label{app:fit-all}
\end{figure}

\begin{figure}
    \centering
    \includegraphics[width=1\linewidth]{  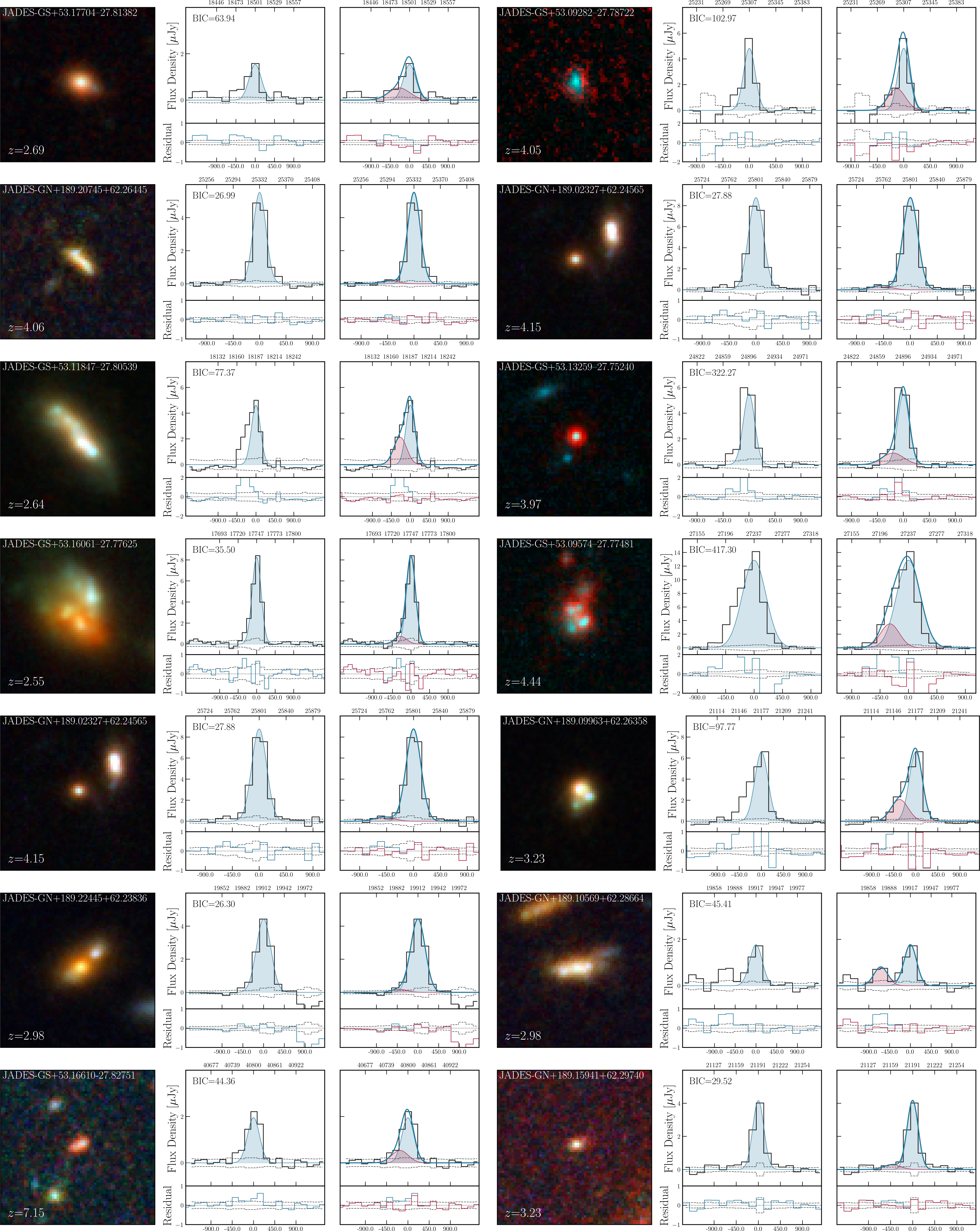}
    \caption{Continuation of Fig. \ref{app:fit-all}}
    \label{app:all_fits_2}
\end{figure}

\begin{figure}
    \centering
    \includegraphics[width=1\linewidth]{  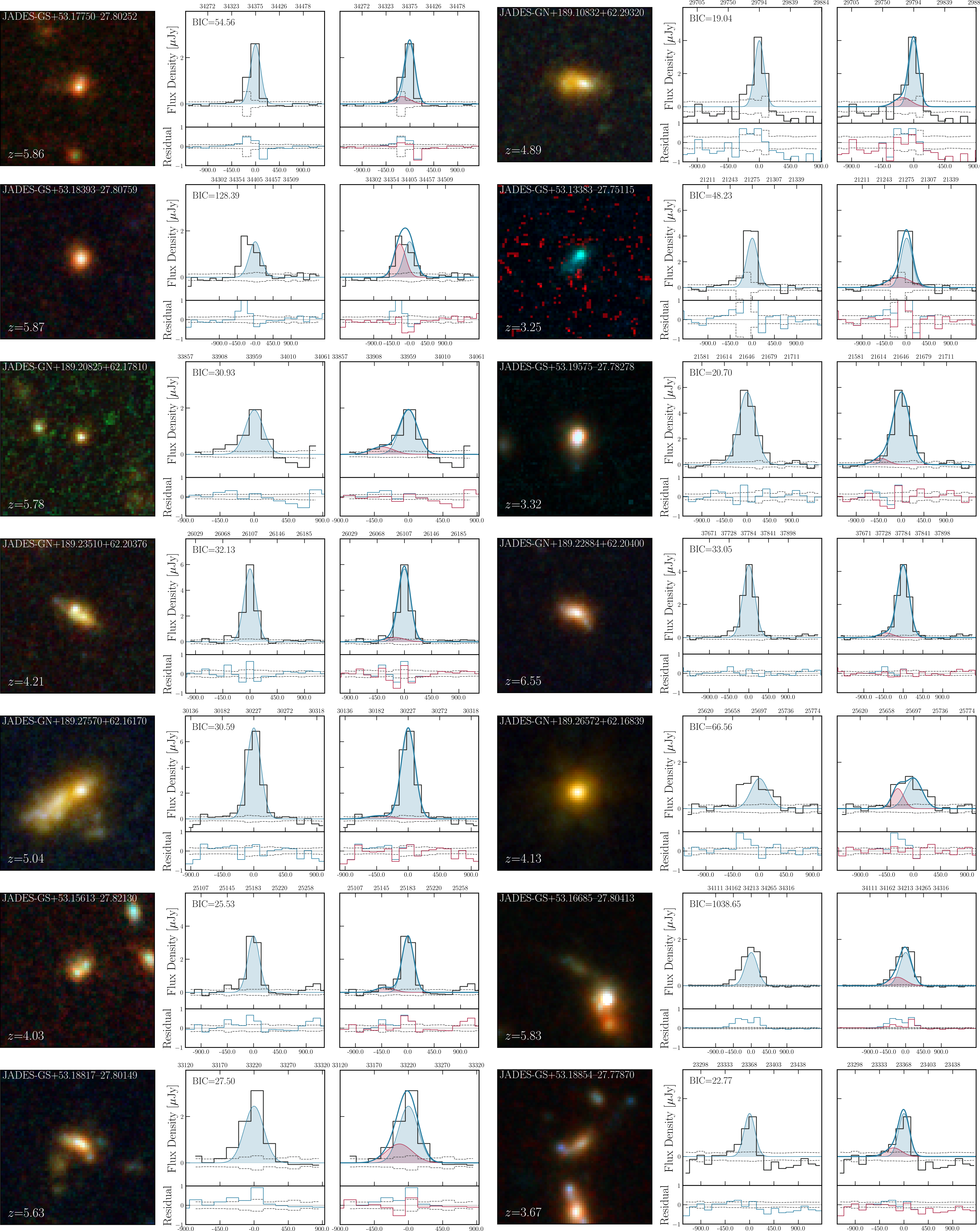}
    \caption{Continuation of Fig. \ref{app:fit-all}}
    \label{app:all_fits_3}
\end{figure}

\bibliography{outflow.bib}{}
\bibliographystyle{aasjournal}

\end{document}